\documentclass[lettersize,journal]{IEEEtran}
\usepackage{array}
\usepackage[caption=false,font=footnotesize]{subfig}
\usepackage{textcomp}
\usepackage{stfloats}
\usepackage{url}
\usepackage{verbatim}
\usepackage{graphicx}
\usepackage{cite}
\usepackage[colorlinks=true, linkcolor=blue, citecolor=blue, urlcolor=blue]{hyperref}
\usepackage{booktabs}
\usepackage{amsmath,amssymb,amsfonts}
\usepackage{graphicx}
\usepackage{textcomp}
\usepackage{xcolor}
\usepackage{subfig}
\usepackage{multirow}
\usepackage{upgreek}
\usepackage{color}
\usepackage{epstopdf}
\usepackage{bm}
\usepackage{array,color}
\usepackage[linesnumbered,ruled,vlined]{algorithm2e}
\usepackage[normalem]{ulem}
\usepackage{amsfonts}
\usepackage{graphics}
\usepackage{epsfig}
\usepackage{soul}
\usepackage{gensymb}
\soulregister\cite7
\soulregister\ref7
\usepackage{setspace}
\usepackage{svg}
\usepackage{tabularx}

\DeclareUnicodeCharacter{2212}{\ensuremath{-}}

\begin{document}
	
\title{Two-Timescale Trajectory Control, Task Reassignment and Resource Scheduling for Multi-UAV Cooperative Mobile Edge Computing}
	
\author{Bin Li, Shiyu Sun, Lei Liu, and Zesong Fei,~\IEEEmembership{Senior Member,~IEEE}

\thanks{Bin Li and Shiyu Sun are with the School of Computer Science, Nanjing University of Information Science and Technology, Nanjing 210044, China (e-mail: bin.li@nuist.edu.cn; 202512201138@nuist.edu.cn).}
\thanks{Lei Liu is with the Guangzhou Institute of Technology, Xidian University, Guangzhou 510555, China (e-mail: leiliu@xidian.edu.cn).}
\thanks{Zesong Fei is with the School of Information and Electronics, Beijing Institute of Technology, Beijing 100081, China (e-mail: feizesong@bit.edu.cn).}
}



\maketitle
	
\begin{abstract}
Uncrewed aerial vehicle (UAV)-assisted mobile edge computing (MEC) is a promising technology, leveraging the advantages of UAVs including flexible deployment, wide coverage, and reliable wireless communication. In this paper, we study the task reassignment and offloading problem in a multi-UAV cooperative operating environment, and consider the real occurrence that the computing tasks arrive more frequently than UAVs make flight decisions. In addition, we also take into account the position offset of the UAV during flight and the unavailability of the UAV due to external interference. We propose a two-timescale optimization framework, where the trajectory of the UAV is carried out on a large timescale, while the task reassignment and computing resource allocation are jointly optimized on a small timescale. The core objective is to maximize the long-term average system gain, which takes both energy consumption and task delay into account. To address the UAV trajectory optimization problem under large timescale constraints, we propose a constrained multi-agent deep reinforcement learning algorithm. For decision-making within small timescale, to jointly optimize task offloading decisions and computing resource allocation, we integrate convex optimization with a heuristic algorithm. Finally, we conduct extensive simulations to evaluate the proposed algorithms against the baseline, achieving a 24\% reduction in latency and a 28\% reduction in total system cost.
\end{abstract}


\section{Introduction}\label{Sec.Introduction}
\IEEEPARstart{T}{he} traditional cloud computing paradigm centralizes all computational tasks in remote data centers for processing, which not only leads to unacceptable transmission delays but also consumes a large amount of core network bandwidth, becoming a key bottleneck restricting the large-scale deployment of the applications. In response to this challenge, mobile edge computing (MEC) has emerged as a revolutionary computing paradigm. By deploying computing and storage resources at the edge of the radio access network, MEC offloads computing capabilities from the cloud to the vicinity of end users, enabling computational tasks to be processed much closer to where data is generated. This architecture can significantly shorten the data transmission distance, reduce service delay and network congestion, while protecting user data privacy, and has become one of the core technologies of fifth-generation and beyond networks \cite{sarah2024drl}. However, the traditional ground MEC systems rely on fixedly deployed base stations and edge servers, whose coverage and service capabilities are strictly limited by geographical environments and infrastructure conditions. In scenarios lacking fixed communication infrastructure, such as natural disasters and temporary gatherings, ground MEC systems often fail to operate normally, even in urban environments with well-developed infrastructure, building obstructions and uneven user distribution can lead to coverage blind spots and computational resource overload in some areas \cite{kuang2023joint,zhang2024energy}.

A novel solution to the above problems is provided by the rapid development of uncrewed aerial vehicle (UAV) technology. UAVs offer distinctive benefits, including deployment flexibility, high maneuverability, broad coverage, and a high likelihood of line-of-sight (LoS) communication, enabling them to act as aerial mobile edge servers that provide on-demand computation offloading for terrestrial users \cite{chen2025joint}. UAV-assisted MEC technology combines the maneuverability advantages of UAVs and the low-latency characteristics of edge computing, enabling dynamic position adjustment according to the real-time distribution of users to achieve precise delivery of computational resources and significantly improve the overall system performance. In the past few years, this field has become a research hotspot in both academia and industry, attracting extensive attention from scholars \cite{zhang2025secure}.

Early works on UAV-assisted MEC task offloading primarily concentrate on single-UAV settings, where system energy consumption or task latency is reduced by jointly optimizing the UAV trajectory, task offloading decisions, and computing resource allocation \cite{qiu2025tom}. However, in complex scenarios such as multi-UAV cooperative operations, a single UAV has limited computing capacity and coverage area, which makes it difficult to address the simultaneous computing requests from a vast number of users \cite{guo2025cognitive}. Multi-UAV collaborative work can form a distributed computing network, enabling the sharing of computing resources and load balancing, thereby providing more powerful and reliable edge computing services \cite{zhou2025novel}. Although some studies have begun to address the multi-UAV cooperative task offloading problem, most of them assume that UAV flight decisions and task offloading decisions are updated on the same time scale \cite{sun2024two}. In practical systems, the arrival frequency of computing tasks is typically much higher than the frequency at which UAVs make flight decisions. Because task arrivals may occur at the millisecond level, while UAV position adjustments are constrained by physical motion laws and energy consumption, and usually only be performed on the time scale of seconds or even minutes \cite{wang2026two}. Ignoring this time-scale disparity not only leads to a sharp increase in algorithmic computational complexity but may also result in frequent and unnecessary UAV maneuvers, causing significant additional energy consumption.

Furthermore, deep reinforcement learning (DRL)-based task offloading algorithms are primarily modeled using standard Markov decision processes (MDPs), which struggle to effectively handle long-term average constraint problems. In practical systems, there are many constraints that must always be satisfied, otherwise task processing will suffer. The standard DRL algorithm can only approximate the handling of constraints by introducing penalty terms into the reward function, and it cannot strictly guarantee the satisfaction of the constraints. This is unacceptable in military and industrial applications with extremely high reliability requirements \cite{bi2021lyapunov,long2025lyapunov}.

To resolve the challenges outlined above, this paper investigates the joint task reassignment and resource allocation problem in multi-UAV collaborative MEC environments. Motivated by the inherent time-scale mismatch between computation task arrivals and UAV flight decisions, as well as the practical uncertainties of UAV flight errors and potential UAV failures, we propose a novel two-timescale optimization framework. Within this framework, UAV flight trajectories are planned on a large time scale to determine optimal service positions for ground users, while task reassignment decisions and computing resource allocation are jointly optimized on a small time scale to rapidly respond to the dynamic arrival of computation tasks. Our objective is to maximize the long-term average system utility, which is defined as a weighted combination of energy consumption and task latency, subject to UAV computing resource constraints and motion constraints.
The main contributions of this paper are summarized as follows:
\begin{enumerate}
	\item In the system model, we enable the UAVs in user-dense areas and the UAVs in user-sparse areas to transfer tasksby means of high-speed communication links between them, thereby reasonably allocating tasks. This not only alleviates the resource shortage of UAVs in user-dense areas but also makes full use of the computing resources of UAVs in user-sparse areas.
	
	\item Different from the common UAV flights, we consider the uncertainty of the final trajectory caused by external interference during the flight of the UAV, and optimize the flight trajectory of the UAV in large time slots. To ensure the constraints that the UAV is subject to, we model the problem as a constrained Markov decision process (CMDP), and propose a constrained multi-agent deep reinforcement learning algorithm to solve this problem.
	
	\item Considering the case where the UAV fails at a certain time, we optimize the task reassignment and computing resource allocation in small time slots, in order to obtain the optimal policy, we adopt a heuristic algorithm combined with convex optimization to solve this problem.
	
\end{enumerate}

The remainder of this paper is organized as follows. Related work is reviewed in Section \ref{Sec.RelatedWork}. The proposed system model and problem formulation are introduced in Section \ref{Sec.Systemmodel}. The two-timescale optimization method is presented in Section \ref{Sec.Algorithm}. Simulation results are provided in Section \ref{Sec.Results}, and the paper is finally summarized in Section \ref{Sec.Conclusion}.

\section{Related Work}\label{Sec.RelatedWork}

UAVs equipped with MEC servers have gained significant attention as mobile edge platforms due to their high mobility and flexible deployment capabilities. 
Bai \textit{et al.} \cite{11021435} studied a dynamic stochastic computation offloading problem in UAV-assisted maritime MEC networks with random task arrivals and used Lyapunov optimization to transform the long-term stochastic optimization problem into deterministic subproblems within each time slot.
Li \textit{et al.} \cite{li2024uav} studied the fair computation offloading problem in UAV-enabled MEC networks and proposed a deep reinforcement learning approach based on an actor-critic parallel architecture to address this issue. 
Similarly, Wang \textit{et al.} \cite{wang2021deep} proposed a dynamic trajectory control approach based on deep reinforcement learning for UAV-assisted MEC systems, which addresses the inherent inefficiency of conventional fixed or pre-planned UAV trajectories that cannot effectively adapt to time-varying user demands and wireless channel conditions, ultimately enhancing the overall task processing efficiency of the system.
For multi-user UAV-assisted MEC systems, Huang \textit{et al.} \cite{huang2024multiobjective} proposed a multi-objective deep reinforcement learning framework for UAV base station-assisted mobile edge computing networks, addressing the coupled optimization challenge of computation offloading and UAV trajectory control to balance and optimize multiple conflicting system performance indicators.
Ye \textit{et al.} \cite{gan2021trajectory} proposed a sequential optimization approach for UAV trajectory and offloading strategy, aiming to minimize both user and UAV energy consumption.
Focusing on air-ground integrated MEC systems, Chen \textit{et al.} \cite{chen2023energy} developed a distributed online algorithm for energy-efficient task offloading that integrates convex optimization with game theory.
Although these single-UAV works have demonstrated promising results, they are limited in their ability to serve large numbers of users simultaneously due to coverage and computational constraints.

To overcome the limitations of single UAV systems, multi-UAV collaborative MEC networks have been extensively studied. 
Li \textit{et al.} \cite{li2023computation} proposed a knowledge-assisted multiagent reinforcement learning approach for multi-UAV-enabled MEC that incorporates expert knowledge to guide agent exploration and adopts a hierarchical collaboration mechanism for multi-UAV cooperative decision-making, achieving significantly faster algorithm convergence and higher task processing success rate with better system fairness.
In \cite{wang2020multi}, Wang \textit{et al.} presented a multi-agent reinforcement learning framework that integrates an attention mechanism, achieving faster convergence speed and reduced weighted energy consumption compared with baseline approaches.
Zhao \textit{et al.} \cite{zhao2022multi} developed a cooperative multi-agent deep reinforcement learning framework for multi-UAV-assisted mobile edge computing systems, which addresses the task offloading challenge under the constraints of limited UAV computation and energy capacities to minimize the overall system execution delay and energy consumption.
Seid \textit{et al.} \cite{seid2021multi} developed a task offloading and resource allocation algorithm for multi-UAV networks, effectively managing the expansive continuous action space to minimize computation costs.
Shao \textit{et al.} \cite{shao2024deep} proposed a priority experience replay-enhanced multi-agent double-delay deep deterministic policy gradient framework for anti-interference UAV edge computing, which is specifically designed to minimize the weighted sum of system delay and UAV energy consumption under malicious interference attacks.
Li \textit{et al.} \cite{li2023robust} presented a robust computation offloading and trajectory optimization framework for multi-UAV-assisted MEC networks, considering both communication and computation uncertainties. Compared to traditional DRL and heuristic algorithms, their approach achieved higher energy efficiency and stronger robustness.

Building upon these foundational multi-UAV studies, recent research has increasingly adopted UAV swarm architectures or layered processing frameworks to address more complex challenges.
Wu \textit{et al.} \cite{wu2025dynamic} designed a hierarchical task allocation framework for UAV swarms, introducing differentiated exploration strategies and dynamic formation reorganization mechanisms to optimize inter-UAV collaboration patterns and enhance responsiveness to sudden mission changes.
To tackle the uncertainty in task sizes and high mobility of UAVs in low-altitude networks, Jiang \textit{et al.} \cite{jiang2025dro} proposed a layered collaborative architecture integrating UAVs and high-altitude platforms, adopting a decoupled hierarchical optimization approach to jointly optimize computation offloading decisions and UAV flight trajectories.
For latency-sensitive applications requiring ultra-low end-to-end performance, Wang \textit{et al.} \cite{wang2025joint} developed a bi-level joint optimization framework for multi-UAV clusters, enabling fine-grained task splitting and distributed load balancing while coordinating UAV deployment positions to minimize overall system latency.
Talukdar \textit{et al.} \cite{talukdar2025ddpg} modeled joint UAV navigation and data collection as an MDP and applied deep deterministic policy gradient (DDPG) to shorten task completion time and reduce data-retrieval energy. In \cite{talukdar2026mappo}, they extended it to multi-UAV MEC-enabled Internet of Things (IoT) networks with wireless energy transfer, using multi-agent proximal policy optimization (MAPPO) to jointly optimize task quality, offloading and resource scheduling. These two studies validate the capability of actor-critic DRL for continuous control and multi-agent cooperation, yet they remain limited to a single-timescale and enforce reliability only via reward shaping, thereby leaving open issues such as timescale mismatch, strict long-term average constraints, and resilience to position offsets and UAV failures.

In view of prior work, there are currently few studies on the issue of the different dynamics of task arrival and UAV flight decisions in the UAV-assisted MEC network, and the real scenarios of UAV failure have not been taken into account. Against this background, we propose a two-timescale hierarchical optimization framework. Its key advantage lies in elegantly decomposing the original coupled problem into two subproblems, which are solved separately on different timescales using tailored methods: UAV trajectory optimization is performed on a large timescale through a constrained multi-agent deep reinforcement learning algorithm, whereas task reassignment and computational resource allocation are jointly optimized on a small timescale by combining convex optimization with heuristic algorithms. This design closely matches the characteristics of practical applications, namely, tasks arrive frequently while UAV flight decisions are updated at a lower frequency, and it also satisfies the robustness requirements against UAV position deviations and failures.

\section{System Model and Problem Formulation}\label{Sec.Systemmodel}
In this section, we first introduce the proposed system model supporting UAV cooperative task processing, followed by the UAV mobility energy consumption model, communication model, and task reassignment and offloading model. Subsequently, we formulate the multi-objective optimization problem.
\subsection{System Model}

\begin{figure}[t!]
	\centering
	\includegraphics[width=0.5\textwidth]{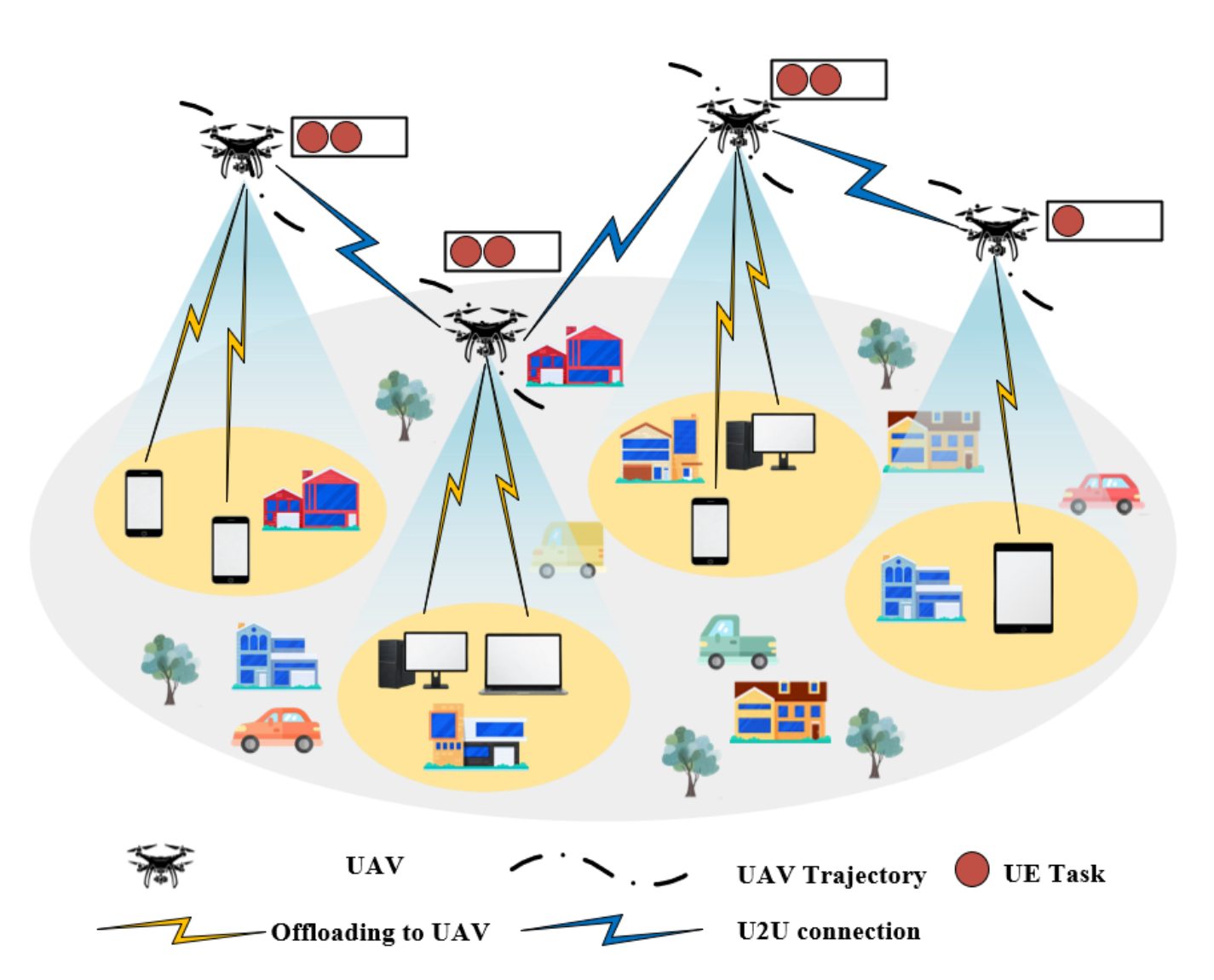}
	\caption{Task offloading system model for the proposed multi-UAV-assisted MEC.}
	\label{fig. system model}
	\vspace{0 cm}
\end{figure}

Fig. \ref{fig. system model} illustrates the system architecture of a multi-UAV-assisted MEC network, we define the collection of indexes for UAVs as $\forall u \in \mathcal{U} \triangleq \{1,2, \ldots,U\}$ and the collection of indexes for user equipment (UE) as $\forall m \in \mathcal{M} \triangleq \{1,2, \ldots,M\}$. UAVs equipped with computational resources are deployed to provide computing services for UEs with typically limited computational capabilities. Owing to the uneven spatial distribution of UEs, UAVs in user-dense areas may be overloaded and suffer from insufficient computing resources and thus increased latency, while UAVs in user-sparse areas may still hold abundant idle capacity. To balance this spatial mismatch between task demand and computing capability, UAVs are able to reassign tasks among themselves via high-speed inter-UAV links, thereby achieving more rational task allocation and better utilization of the computing resources of all UAVs.

Since the computational tasks usually arrive frequently, UAVs make flight decisions on a larger timescale to avoid unnecessary decisions upon each task arrival. Therefore, we consider a two-timescale MEC framework consisting of large-timescale UAV trajectory design and small-timescale task offloading. We define the time period of large time slot as $\Delta$ and the time period of small time slot as $\tau$, where $\Delta = N \tau$. The continuous system time is discretized into frames of equal duration. Thus, we define the set of large time slots as $\mathcal{L} \triangleq \{1,2, \ldots,L\}$, and the set of small time slots as $\mathcal{T} \triangleq \{1,2, \ldots,T\}$. In each small time slot, each UE generates a task. The task generated by UE $m$ in time slot $t, t \in \mathcal{T}$ is characterized by $\langle D_m(t), C_m(t)\rangle$, where $D_m(t)$ is the task size in bits and $C_m(t)$ is the computational intensity of the task in cycles per bit.
The wireless communication channels between UEs and UAVs, as well as the communication links between UAVs, are modeled as time-varying. Although the channel conditions fluctuate at the beginning of each time slot, they are assumed to remain constant throughout the entire time slot. This quasi-static assumption is reasonable, as each time slot $\tau$ has a very short duration under typical conditions.

\subsection{MEC Mobility Energy Consumption Model}
A three-dimensional (3D) Cartesian coordinate system is considered without loss of generality. UEs are located on the 2D ground plane, and their positions are denoted by the set $\mathcal{W}_{\mathcal{M}} = \{\mathbf{w}_1, \mathbf{w}_2, \dots, \mathbf{w}_M\}$, where $\mathbf{w}_m = [x_m, y_m, 0]$ for $m \in \mathcal{M}$. In modern IoT applications such as disaster response and precision agriculture, UAVs are equipped with advanced sensors and communication capabilities that enable them to accurately estimate or retrieve the positions of UEs. Therefore, it is assumed that the UEs stay stationary and their positions can be fetched by the UAV network in real time. The set of 3D positions of UAVs is denoted by $\mathcal{Q}_{\mathcal{U}}[l] = \{\mathbf{q}_1[l], \mathbf{q}_2[l], \dots, \mathbf{q}_U[l]\}$, where the position of each UAV at time $l$ in the large time slot is $\mathbf{q}_u[l] = [x_u, y_u, z_u] \in \mathbb{R}^3$ and its velocity is $\mathbf{v}_u[l] = [v_x^u[l], v_y^u[l], v_z^u[l]]$ for $u \in \mathcal{U}$. Let $V_{\text{max}}$ be the maximum velocity of UAVs, then the maximum distance each UAV can travel between two consecutive large time slots is $D_{\text{max}} = V_{\text{max}} \Delta$. We also define the flight area of the UAV $\mathcal{D} \in \mathbb{R}^3$. In addition, collision avoidance requires that UAVs keep a safe distance from each other; thus, for any pair of UAVs, the separation must be greater than the minimum safe distance $D_{\text{min}}$ at any large time slot. Therefore, the UAV mobility limitations are modeled as \cite{zeng2017energy}
\begin{subequations}
\begin{align}
	\hat{\mathbf{q}}_u[l+1] &= \mathbf{q}_u[l] + \mathbf{v}_u[l] \Delta, ~ \forall u \in \mathcal{U}, l \in \mathcal{L}, \label{eq:uav:dyn} \\
	\|\mathbf{v}_u[l]\| &\leq V_{\text{max}}, ~ \forall u \in \mathcal{U}, l \in \mathcal{L}, \label{eq:uav:vel} \\
	\|\mathbf{q}_u[l] - \mathbf{q}_i[l]\| &\geq D_{\text{min}}, ~ \forall u \neq i \in \mathcal{U}, l \in \mathcal{L}, \label{eq:uav:col} \\
	\|\mathbf{q}_u[l+1] - \mathbf{q}_u[l]\| &\leq D_{\text{max}} = V_{\text{max}} \Delta, ~ \forall u \in \mathcal{U}, l \in \mathcal{L}, \label{eq:uav:step}
\end{align}
\end{subequations}
where (\ref{eq:uav:dyn}) represents the position update of UAVs and $\hat{\mathbf{q}}_u[l+1]$ represents the position that the UAV should reach without external interference, (\ref{eq:uav:vel}) denotes the velocity constraint, (\ref{eq:uav:col}) guarantees a safe separation between any two UAVs to prevent collisions, and (\ref{eq:uav:step}) imposes a restriction on the maximum flight distance allowed during each large time slot.

The propulsion energy consumption of UAV $u$ flying at velocity $\mathbf{v}_u[l]$ is given by \cite{zeng2019energy}:
\begin{equation}
	\begin{aligned}
		E_u^{\text{fly}}[l] = & \left( P_0 \left(1 + \frac{3\|\mathbf{v}_u[l]\|^2}{U_{\text{tip}}^2} \right) + \right. \\ 
		& P_{\text{ind}} \left( \sqrt{1 + \frac{\|\mathbf{v}_u[l]\|^4}{4v_0^4}} - \frac{\|\mathbf{v}_u[l]\|^2}{2v_0^2} \right)^{1/2} + \\
		& \left. \frac{1}{2} d_0 \rho_0 s A \|\mathbf{v}_u[l]\|^3 + Wg \left\| v_z^u[l] \right\|\right) \Delta, 
	\end{aligned}
\end{equation}
where
$P_0$ is the blade profile power,
$P_{\text{ind}}$ is the induced power,
$U_{\text{tip}}$ is the blade tip speed of the rotor,
$v_0$ denotes the average rotor induced velocity in hover,
$d_0$ denotes the fuselage drag coefficient,
$\rho_0$ denotes the air density,
$s$ denotes the rotor solidity (the ratio of the total blade area to the rotor disk area),
$A$ indicates the area swept by the UAV rotor blades,
$W$ represents the weight of the UAV and $g$ represents the gravitational acceleration.

The total energy consumption of the system is given by:
\begin{equation}
E_{\text{total}} = \sum_{l=1}^{L} \sum_{u=1}^{U} E_u^{\text{fly}}[l].
\end{equation}

\subsection{Channel Model}
This section presents the channel model adopted in the UAV-based access MEC network. We consider a UAV-augmented MEC system involving cooperation among multiple UAVs that can communicate with each other. In this system, orthogonal frequency-division multiple access is adopted as the multiple access scheme to coordinate the uplink/downlink transmissions between UAVs and UEs. We focus on describing two types of communication links, including air-to-ground (A2G) and air-to-air (A2A). The A2G channels between UAVs and UEs are affected by both LoS and non-line-of-sight (NLoS) propagation conditions, depending on environmental characteristics and UAV flight altitude. The A2A link between UAVs is typically an unobstructed LoS path due to high-altitude operation. In addition, the cross-interference between UAVs and UEs is out of the scope of this paper and can be further explored in the future.

\subsubsection{A2G Path Loss Model}
In order to precisely characterize the path loss between $u \in \mathcal{U}$ and ground UE $m \in \mathcal{M}$, we observe that A2G communication links encounter many scatterers and obstacles in real-world environments. Obstacles cause shadowing and scattering, preventing radio signals from propagating in free space and resulting in additional path loss. Consequently, the simplified free-space path loss (FSPL) model cannot adequately describe the communication between ground UEs and UAVs. To address this, a probabilistic path loss model is employed for A2G channels, one that incorporates the likelihood of LoS and NLoS conditions as well as the corresponding path losses. In this paper, we adopt a probabilistic approach to account for both LoS and NLoS conditions simultaneously. The LoS connection probability $P_{\text{LoS}}$ is determined by the elevation angle and environmental parameters, and is expressed as \cite{al2014optimal}
\begin{equation}
P_{\text{LoS}} = \frac{1}{1 + \vartheta \exp \left(-\xi \left(\phi_{u,m} - \vartheta \right) \right)},
\end{equation}
where $\vartheta$ and $\xi$ are environment-related constants, and $\phi_{u,m}$ is the elevation angle from UE $m$ to UAV $u$, given by
\begin{equation}
\phi_{u,m} = \frac{180}{\pi} \arctan\left( \frac{z_u}{\sqrt{(x_u - x_m)^2 + (y_u - y_m)^2}} \right).
\end{equation}
The NLoS probability is
\begin{equation}
P_{\text{NLoS}} = 1 - P_{\text{LoS}}.
\end{equation}
The path loss under LoS and NLoS conditions is given by
\begin{equation}
	\begin{aligned}
		PL_{u,m}^{\text{LoS}} =20 \log _{10}\left(d_{u,m}\right)+20 \log _{10}\left(f_{c}\right)\\
		+20 \log _{10}\left(\frac{4 \pi}{c}\right)+\eta_{\text{LoS}},
	\end{aligned}
\end{equation}
\begin{equation}
	\begin{aligned}
	PL_{u,m}^{\text{NLoS}} =20 \log _{10}\left(d_{u,m}\right)+20 \log _{10}\left(f_{c}\right)\\
	+20 \log _{10}\left(\frac{4 \pi}{c}\right)+\eta_{\mathrm{NLoS}},
	\end{aligned}
\end{equation}
where $d_{u,m}$ is the 3D distance between UAV $u$ and UE $m$, $c$ is the speed of light, 
$f_c$ denotes the carrier frequency, $\eta_{\text{LoS}}$ and $\eta_{\text{NLoS}}$ indicate the additional average losses under LoS and NLoS conditions.
The average path loss is then
\begin{equation}
PL_{u,m} = P_{\text{LoS}} \cdot PL_{u,m}^{\text{LoS}} + P_{\text{NLoS}} \cdot PL_{u,m}^{\text{NLoS}}.
\end{equation}

\subsubsection{A2A Path Loss Model}
The A2A communication between UAVs is dominated by LoS propagation, and thus the FSPL model is adopted \cite{friis1946note}
\begin{equation}
\text{FSPL}_{u,u'} = 20 \log_{10}\left( \frac{4 \pi f_a d_{u,u'}}{c} \right),
\end{equation}
where $f_a$ is the carrier frequency of the A2A channel, and $d_{u,u'}$ is the 3D distance between UAV $u$ and UAV $u'$.

\subsection{MEC Task Reassignment and Offloading Model}
To offload a computational task, UE $m$ must upload the task to the UAV $u$ that ultimately executes it through an accessible UAV $u'$. In each small time slot $t$, UE $m$ may be within the coverage area of multiple UAVs. Among these UAVs, we adopt the widely used proximity-based UAV association scheme, which assigns user tasks to the nearest UAV. Let $x_{m,u}^l$ denote the association indicator between UAV $u$ and UE $m$ in time slot $l$, where $x_{m,u}^l = 1$ if user $m$ is connected to UAV $u$, and $x_{m,u}^l = 0$ otherwise. Formally, the association indicator is determined as
\begin{equation}
x_{m,u}^l = \underset{u \in \mathcal{U}}{\arg\min} \left\| \mathbf{w}_m - \mathbf{q}_u[l] \right\|,
\end{equation}
with the constraint
\begin{equation}
\sum_{u \in \mathcal{U}} x_{m,u}^l = 1, \quad \forall m \in \mathcal{M}, l \in \mathcal{L},
\end{equation}
which ensures that each UE is connected to exactly one UAV at any given time.

There are significant performance drawbacks if only the static scheduling strategy of assigning tasks to the nearest UAV is used. On the one hand, the UAV will move across large time slots during the task execution, resulting in the edge users who were originally in the service range gradually losing high-quality connections. On the other hand, in user-dense areas, the limited number of UAVs and their computing resources are difficult to meet the task processing needs for a large number of UEs at the same time, resulting in a sharp increase in service delay. In this context, task reassignment can be understood as offloading the task to a different UAV to ensure optimal performance. We define a variable $\gamma_{m,u}^t$ to indicate the task service status, whether UAV $u$ is serving the task of UE $m$ in small time slot $t$. Specifically, $\gamma_{m,u}^t = 1$ if UAV $u$ is serving the task of UE $m$, and $\gamma_{m,u}^t = 0$ otherwise. At any given time, each task can only be processed by one UAV, ensuring
\begin{equation}
\sum_{u \in \mathcal{U}} \gamma_{m,u}^t = 1, \quad \forall m \in \mathcal{M}, t \in \mathcal{T}.
\end{equation}
The task processing capability of a UAV is limited by its computational capacity, as defined by the following constraint
\begin{equation}
\sum_{m \in \mathcal{M}} \gamma_{m,u}^t f_{m,u}^t \leq \varpi_u, \quad \forall u \in \mathcal{U}, t \in \mathcal{T},
\end{equation}
where $f_{m,u}^t$ is the central processing unit computational resource allocated by UAV $u$ to UE $m$ at time slot $t$, and $\varpi_u$ represents all the computing resources owned by UAV $u$. This constraint guarantees that the cumulative resource requirement of all served tasks does not exceed the available resources of the UAV.

The A2G transmission latency of user $m$ to UAV $u'$ in time slot $t$ is
\begin{equation}
T_{m,u'}^{\text{A2G}} = \frac{D_m(t)}{r_{u',m}},
\end{equation}
where the A2G link rate is
\begin{equation}
r_{u',m} = B \log_{2}\left( 1 + \frac{P_{m}^t 10^{-\text{PL}_{u',m}/10}}{\sigma^2} \right),
\end{equation}
and $B$ is the link bandwidth, $P_{m}^t$ is the transmit power of the UE $m$, and $\sigma^2 = N_0 B$ is the noise power.
The A2A transmission latency between UAV $u'$ and UAV $u$ is
\begin{equation}
T_{u',u}^{\text{A2A}} = \frac{D_m(t)}{r_{u,u'}},
\end{equation}
where the A2A link rate is
\begin{equation}
r_{u,u'} = B \log_{2}\left( 1 + \frac{P_{u'}^t 10^{-\text{FSPL}_{u,u'}/10}}{\sigma^2} \right).
\end{equation}
The computation execution latency of the task on UAV $u$ is
\begin{equation}
T_{u}^{\text{C}} = \frac{D_m(t) C_m(t)}{f_{m,u}^t}.
\end{equation}
The total latency for UE $m$ to execute the task within time period $t$ is
\begin{equation}
T_m(t) = \sum_{u' \in \mathcal{U}} (x_{m,u'}^t T_{m,u'}^{\text{A2G}} + \sum_{u \in \mathcal{U}}( \gamma_{m,u}^t T_{u',u}^{\text{A2A}} + T_{u}^{\text{C}})).
\end{equation}
The total latency of the system is
\begin{equation}
T_{\text{total}} = \sum_{t=1}^{T} \sum_{m=1}^{M} T_m(t).
\end{equation}

\subsection{Problem Formulation}
In this section, we formulate the joint optimization problem of task assignment and UAV trajectory in a multi-UAV-assisted MEC network. Our objective is to minimize the total delay and UAV energy consumption by optimizing the task offloading decisions $\mathbf{A} = \{\gamma_{m,u}^t, \forall m,u,t\}$, computational resource allocation $\mathbf{F} = \{f_{m,u}^t, \forall m,u,t\}$, and UAV trajectory control $\mathbf{V} = \{\mathbf{v}_u[l], \forall u,l\}$. Therefore, the multi-objective optimization problem can be formulated as
\begin{subequations}
\begin{align}
	\mathbf{P}: \quad & \min_{\mathbf{A}, \mathbf{F}, \mathbf{V}} \quad \left\{ T_{\text{total}}, \, E_{\text{total}} \right\}  \\
	\text{s.t.} \quad & x_{m,u}^t \in \{0, 1\}, ~ \forall u \in \mathcal{U}, m \in \mathcal{M}, t \in \mathcal{T}, \label{eq:constraint:assignment:binary} \\
	& \gamma_{m,u}^t \in \{0, 1\}, ~ \forall u \in \mathcal{U}, m \in \mathcal{M}, t \in \mathcal{T}, \label{eq:constraint:computation:binary} \\
	& \gamma_{m,u}^t = 0, ~ u = \mathcal{X}(t), \label{eq:constraint:uav:no:local} \\
	& \sum_{u \in \mathcal{U}} x_{m,u}^t = 1, ~ \forall m \in \mathcal{M}, t \in \mathcal{T}, \label{eq:constraint:assignment:exclusive} \\
	& \sum_{u \in \mathcal{U}} \gamma_{m,u}^t = 1, ~ \forall m \in \mathcal{M}, t \in \mathcal{T}, \label{eq:constraint:computation:exclusive} \\
	& \sum_{m \in \mathcal{M}} \gamma_{m,u}^t f_{m,u}^t \leq \varpi_u, ~ \forall u \in \mathcal{U}, t \in \mathcal{T},\label{eq:constraint:cpu:capacity}\\
	& f_{m,u}^t \leq \varpi_u, ~ \forall u \in \mathcal{U}, m \in \mathcal{M}, t \in \mathcal{T}, \label{eq:constraint:task:cpu:limit} \\
	& \mathbf{q}_u[l] = \hat{\mathbf{q}}_u[l] + \Delta \mathbf{q}_u[l], ~ \forall u \in \mathcal{U}, l \in \mathcal{L},\label{eq:constraint:position:measurement} \\
	& \Delta \mathbf{q}_u[l] \sim \mathcal{N}\left(0, \epsilon_0^2 \boldsymbol{I}\right), ~ \forall u \in \mathcal{U}, l \in \mathcal{L},\label{eq:constraint:position:noise} \\
	& T_m(t) \leq \tau, ~ \forall m \in \mathcal{M}, l \in \mathcal{L}, t \in \mathcal{T},  \label{eq:constraint:delay:deadline}\\
	& \mathbf{q}_u[l] \in \mathcal{D}, ~ \forall u \in \mathcal{U}, l \in \mathcal{L} \label{eq:constraint:area} \\
	& (\ref{eq:uav:dyn}) - (\ref{eq:uav:step}) \label{eq:constraint:uav:dynamics},
\end{align}
\end{subequations}
where constraint (\ref{eq:constraint:uav:no:local}) defines the set of unavailable UAVs at each small time slot. Specifically, each UAV independently becomes unavailable with probability $p$ at each small time slot, following an i.i.d. Bernoulli process; the availability state is re-drawn independently at every slot, hence a UAV that is available at the next slot can resume normal task processing. The resulting outage duration is therefore geometrically distributed with mean $1/p$, which characterizes transient, recoverable unavailability rather than permanent failure. A UAV that is unavailable at slot $t$ cannot receive, compute, or forward any task during that slot. Constraints (\ref{eq:constraint:position:measurement}) and (\ref{eq:constraint:position:noise}) consider the position offset of UAVs during flight. Constraint (\ref{eq:constraint:delay:deadline}) ensures that the processing latency of user tasks is within a controllable range. In addition, constraint (\ref{eq:constraint:uav:dynamics}) represents the mobility constraints of UAVs.

Problem $\mathbf{P}$ involves both binary variables and continuous variables, while the expressions in (\ref{eq:uav:col}) and (\ref{eq:constraint:computation:exclusive}) are non-convex constraints. Consequently, we find that problem $\mathbf{P}$ is a mixed-integer non-linear programming problem \cite{shi2022joint}, and it cannot be efficiently solved in polynomial time using traditional methods such as the simplex or interior-point techniques. In particular, when there is no prior information about the dynamic channel and network context, traditional offline algorithms are incapable of making real-time decisions to find a solution. This is due to the fact that offline optimization needs the complete network state beforehand. Therefore, conventional iterative offline methods are hard to apply for timely problem resolution. As a machine learning approach, multi-agent DRL (MADRL) can interact with and learn from the environment, ultimately obtaining a device-deployable policy model that facilitates real-time decision-making based on the current state and achieves long-term benefits.

\section{Two-Timescale Optimization Framework}\label{Sec.Algorithm}
In this section, we first introduce the joint optimization problem of UAV trajectory control and task offloading based on the two-timescale optimization model, and elaborate on our proposed hierarchical optimization framework.

We adopt a two-timescale optimization approach that decomposes the joint UAV trajectory control and task offloading problem into a large-timescale subproblem (\ref{eq:large}) and a short-timescale subproblem (\ref{eq:short}). The large-timescale subproblem focuses on optimizing UAV trajectories by determining their flight velocities to minimize the overall system latency under controllable UAV flight energy consumption. From the UAV positions obtained from the large-timescale optimization, the short-timescale subproblem selects the optimal offloading strategy and allocates UAV computational resources according to the sizes of user tasks. By jointly solving the large-timescale and short-timescale subproblems, we achieve balanced task offloading that harmonizes computational load distribution and UAV trajectory control. The two-timescale optimization problem is formulated as

\textbf{Long-Timescale:}
\begin{subequations}
\begin{align}
	\mathbf{P1}: \quad & \min_{\mathbf{V}} \quad \left\{ T_{\text{total}}, \, E_{\text{total}} \right\} \label{eq:large}\\
	&\text{s.t.} \quad (\ref{eq:uav:dyn}) - (\ref{eq:uav:step}), (\ref{eq:constraint:area}).
\end{align}
\end{subequations}

\textbf{Short-Timescale:}
\begin{subequations}
\begin{align}
	\mathbf{P2}: \quad & \min_{\mathbf{A}, \mathbf{F}} \quad T_{\text{total}} \label{eq:short}\\
	&\text{s.t.} \quad \quad (\ref{eq:constraint:assignment:binary}) - (\ref{eq:constraint:delay:deadline}).
\end{align}
\end{subequations}

\begin{figure*}[t!]
	\centering
	\includegraphics[width=\textwidth]{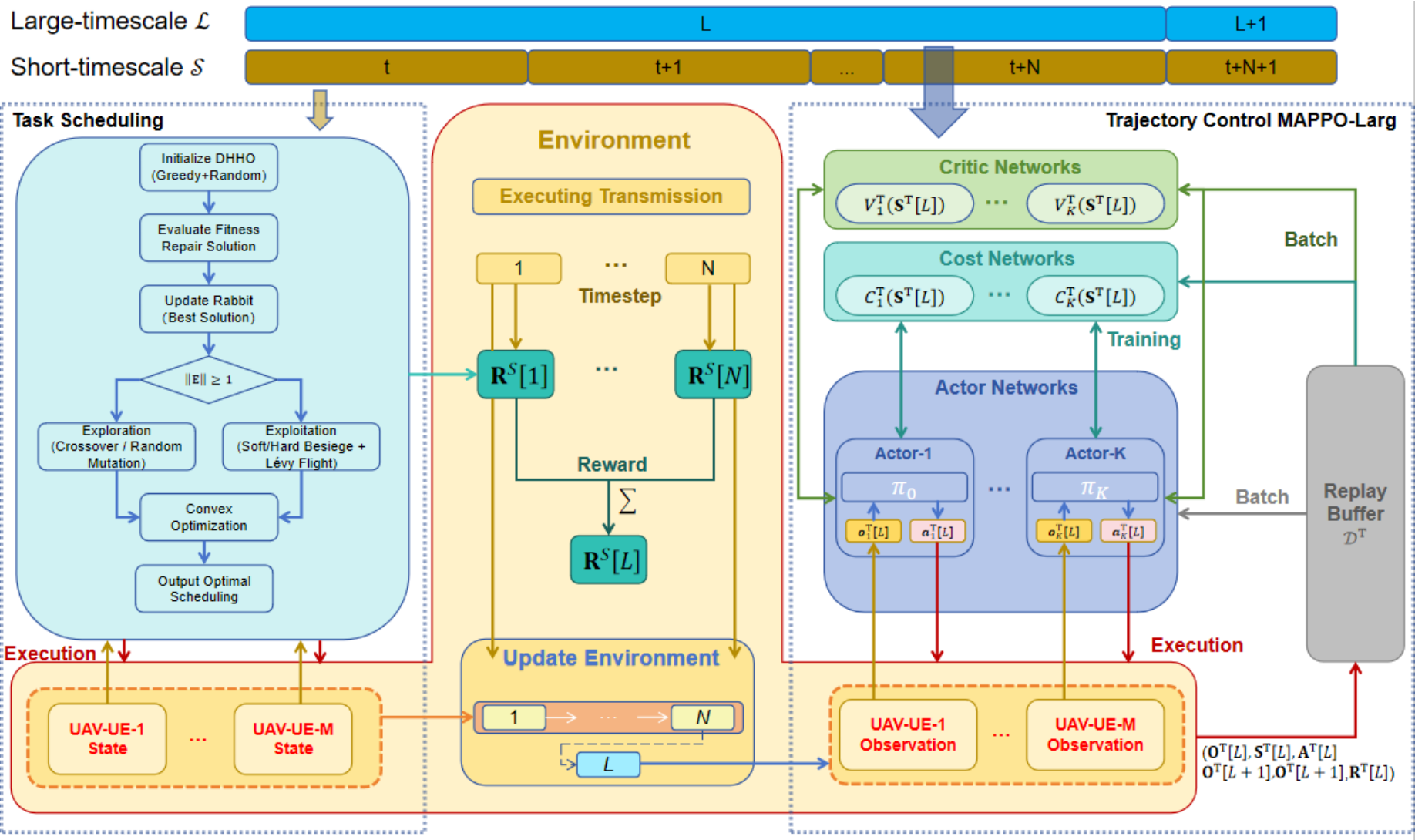}
	\caption{Overall structure and workflow of the proposed algorithm.}
	\label{fig:algorithm-model}
	\vspace{0 cm}
\end{figure*}

\subsection{Large-Timescale Optimization}
Directly solving problem $\mathbf{P1}$ still poses several challenges. First, multi-UAV-assisted MEC scenarios exhibit several realistic characteristics, which complicates the design of low-complexity and efficient algorithms that can not only needs to meet the flight constraints of the UAV, but also needs to take into account the energy consumption of the UAV, ensure the delay of the user task, and adapt to the dynamic characteristics of the system. Furthermore, optimizing the objectives of problem $\mathbf{P1}$ often leads to conflicting results. For example, reducing UAV flight energy consumption may force UAVs to remain in positions with poor communication conditions, resulting in increased latency and reduced number of offloaded tasks.

To address this challenge, we utilize the linear weighting method, one of the most widely used multi-objective optimization approaches \cite{stanimirovic2011linear,alsyouf2017multi}, due to its low algorithmic complexity and ability to quickly produce effective solutions. Additionally, it allows us to effectively balance the importance of each goal and obtain satisfactory solutions within the constraints of our problem. Therefore, we transform the multi-objective optimization problem into a single-objective optimization problem as
$$
\rho(\mathbf{V}) = w_1 \cdot \frac{T_{\text{total}} - T_{\text{MAX}}}{T_{\text{MAX}}} + w_2 \cdot \frac{E_{\text{total}} - E_{\text{MAX}}}{E_{\text{MAX}}}
$$
where the non-negative parameters $w_1$ and $w_2$ denote the weighting factors for latency and energy consumption, respectively. $T_{\text{MAX}}$ is the worst-case total latency, and $E_{\text{MAX}}$ indicates the energy consumed at the fastest flight speed. Note that the weighting coefficients can be adjusted according to users' preferences for different objectives. At this point, problem $\mathbf{P1}$ is transformed into $\mathbf{P1'}: \min \rho(\mathbf{V})$. However, the problem still remains a complex non-convex problem due to the inclusion of UAV distance constraints. Therefore, we propose a MAPPO-Lagrangian training framework for multi-UAV-assisted MEC networks, which can find the best possible solution under multiple constraints.

Conventional MDP-based optimization approaches usually assume that each agent has full global observations, which makes them unsuitable for dynamic and uncertain multi-agent environments \cite{liu2024two}, and fail to adequately consider the impact of constraints on optimization objectives. Thus, we formulate our problem as a constrained multi-agent Markov
decision process (CMADP) to achieve partial observability of decisions and guarantee the constraints. Under the CMDP framework, the primal-dual algorithms employing the Lagrangian method have been proven to converge to the optimal constrained policy under mild assumptions. Specifically, by reformulating the CMDP as a constrained saddle-point problem, where the primal and dual variables correspond to the minimizing and maximizing players, respectively, the policy gradient primal-dual algorithms with non-asymptotic convergence guarantees can be designed \cite{chen2026primal,ding2020natural}. In general, a CMADP for a set of agents can be represented as $\mathcal{G} = \langle \mathcal{K}, \{\mathcal{O}_k\}_{k \in \mathcal{K}}, \mathcal{S}, \{\mathcal{A}_k\}_{k \in \mathcal{K}},  \mathcal{P}, \gamma$,$\{\mathcal{C}_k\}_{k \in \mathcal{K}}, \{\mathcal{R}_k\}_{k \in \mathcal{K}}$, $\{\pi_k\}_{k \in \mathcal{K}} \rangle$.
Here, $\pi_k(a_k | o_k)$ gives the probability of choosing action $a_k$ given observation $o_k$, while a deterministic policy is usually expressed as $\mu_k(o_k)$, which maps each observation directly to a single action. In multi-agent environments, the global state $\mathbf{S} \in \mathcal{S}$ is only partially observable; consequently, agent $k$ can only perceive a partial observation $o_k \in \mathcal{O}$ from the environment. After taking action $a_k$, the agent receives a scalar reward $r_k$ through the reward function $\mathcal{R}$.

The components of our CMADP formulation are defined as follows:
\begin{enumerate}
	
\item \textit{Agent Set $\mathcal{K}$}: Each UAV acts as an agent that learns flight speed strategies based on its own location and the user's location to minimize the total system cost. Therefore, the agent set is $\mathcal{K} = \{1, \dots, U\}$.

\item \textit{Observation Space $\mathcal{O}_k$}: The observation space of agent $k$ consists of the data that agent $k$ can observe. The observation $o_k(l)$ at large time slot $l$ includes the current position of agent $k$, the positions of the three nearest UAVs to itself $\mathcal{Q}_{k'}[l]$, and the positions of all UEs. Therefore, the local observation is given as 
	\begin{equation}
		o_k(l) = \{\mathbf{q}_k[l], \mathcal{Q}_{k'}[l],\mathcal{W}_{\mathcal{M}}\}.
	\end{equation}

\item \textit{State Space $\mathcal{S}$}: The state space represents all observable data of the entire system. The state $s(l)$ at large time slot $l$ consists of the 3D coordinates of all UAVs and the 2D coordinates of all UEs, which is given by
\begin{equation}
s(l) = \{\mathcal{Q}_{\mathcal{U}}[l], \mathcal{W}_{\mathcal{M}}\}. 
\end{equation}

\item \textit{Action Space $\mathcal{A}_k$}: Each UAV needs to determine its own motion parameters, which are the velocities in the three directions. Therefore, the action $a_k(l)$ of UAV $k$ at large time slot $l$ is expressed as
\begin{equation}
a_k(l) = \{v_x^k[l], v_y^k[l], v_z^k[l]\}.
\end{equation}
Thus, the action space $\mathcal{A}_k$ of UAV $k$ is a continuous set.

\item \textit{Cost Function $\mathcal{C}_k$}: The cost function represents the system penalty imposed on agent $k$ when it violates constraints, which include the UAV speed constraint, the inter-UAV safe distance constraint, and the UAV flight boundary constraint. The cost function of the corresponding small time slot collected by UAV $k$ at large time slot $l$ is defined as
\begin{equation}
c_k(l) = \sum_{t=1}^{N} \left( \eta_1(t) + \eta_2(t) + \eta_3(t) \right)>0,
\end{equation}
where $\eta_1(t)$, $\eta_2(t)$, and $\eta_3(t)$ are the penalties corresponding to UAV violations of velocity constraints, boundary constraints, and safe distance constraints at small time slot $t$, respectively.

\item \textit{Reward Function $\mathcal{R}_k$}: To solve the constructed task offloading optimization problem, $K$ agents need to collaboratively minimize the total system cost while satisfying various constraints. Therefore, the reward function $\mathcal{R}_k(l)$ of UAV $k$ is defined as follows: if all constraints are satisfied, the reward is set to the system cost; if any constraint condition is violated, the relevant penalty terms are appended to the reward function. Based on the above considerations, the reward function of UAV $k$ is defined as
\begin{equation}
\begin{aligned}
r_k(l) = & w_1 \left(\sum_{t=1}^{N} \sum_{m=1}^{M} \gamma_{m,k}^t \frac{T_{\text{MAX}}-T_m(t)}{T_{\text{MAX}}} \right) \\
& + w_2 \left( \frac{ E_{\text{MAX}}-E_k^{\text{fly}}[l]}{E_{\text{MAX}}} \right) - c_k(l).
\end{aligned}
\end{equation}
\end{enumerate}

On-policy DRL approaches are recognized for their training stability and computational efficiency, allowing devices to reserve more resources for other vital functions. Therefore, the paper presents MAPPO-Lagrangian, a refined algorithm for training multi-agent policies that achieves outstanding results on the target task and maintains stable training. As an on-policy MADRL approach based on the actor-critic paradigm, MAPPO-Lagrangian has exhibited excellent performance on diverse tasks. In MAPPO-Lagrangian, the actor network $\theta_k$ outputs actions, the critic network $\omega_k$ estimates the state value function, and the cost network $\varphi_k$ estimates the state cost function. The policy of UAV agent $k$ is denoted by $\pi_{\theta_k}$.

We first present the complete system training framework, in order to facilitate the deployment in a distributed network, the centralized training and decentralized execution framework shown on the right of Fig. \ref{fig:algorithm-model} is adopted. In this framework, each UAV maintains its own experience replay buffer along with a reward critic network and a cost critic network. Flight control is executed using actions generated by the local actor network, and the resulting rewards and constraint-violation penalties are stored in the agent's individual buffer. A centralized training center constructs the global environment state by aggregating the observations collected from all agents, then employs this global state to obtain predicted values through the reward and cost critics. Based on these centralized value estimates, the parameters of each agent's actor, reward critic, and cost critic networks are updated. After the updates are completed, only the parameters of the actor networks are downloaded to the UAVs for subsequent decentralized execution.

In this framework, the state value function of agent $k$ is expressed as
\begin{equation}
\begin{aligned}
&V_{k}(s(n)) = \\
&\quad \mathbb{E} \left\{ \sum_{l=0}^{\infty} \gamma^l \mathcal{R}_k \left( s(n+l), a_k(n+l) \mid s(n) = s \right) \right\},
\end{aligned}
\end{equation}
where $\mathbb{E}\{\cdot\}$ denotes the expectation operation, $\mathcal{R}_k$ is the reward function of agent $k$, and the discount factor $\gamma$ represents the importance of future rewards to all agents.
The action value function can be expressed as
\begin{equation}
\begin{aligned}
&Q_k(s(n), a_k(n)) = \\
&\quad \mathbb{E} \left\{ \sum_{l=0}^{\infty} \gamma^l \mathcal{R}_k \left( s(n+l), a_k(n+l) \right) \mid s(n) = s, a_k(n) = a \right\}.
\end{aligned}
\end{equation}
To obtain the advantage of each action that can be utilized in policy updating, the critic advantage function is given by
\begin{equation}
A_k(s(n),a_k(n)) = Q_k(s(n), a_k(n)) - V_{k}(s(n)),
\end{equation}
which can be estimated using the state value $V_{k}(s(n))$ as
\begin{equation}
\begin{aligned}
&\hat{A}_k(s(n)) = \\
&\sum_{l=0}^{\infty} (\gamma \lambda)^l \left( r_k(n+l) + \gamma V_{k}(s(n+l+1)) - V_{k}(s(n+l)) \right).
\end{aligned}
\end{equation}

It is worth mentioning that the critic advantage function is estimated via generalized advantage estimation, and the parameter $\lambda$ therein serves to balance the bias and variance of the obtained rewards. In addition, $\delta(n+l) = r_k(n+l) + \gamma V_{k}(s(n+l+1)) - V_{k}(s(n+l))$ denotes the temporal difference error. Similarly, $C_k(s(n))$ can be used to calculate the cost advantage function, resulting in $\hat{A}_k^C(s(n))$. Finally, the overall advantage function of agent $k$ is expressed as
\begin{equation}
	\hat{M}_k(s(n)) = \hat{A}_k(s(n)) - \epsilon_k \hat{A}_k^C(s(n)),
\end{equation}
where $\epsilon_k$ is used to determine the degree to which the loss advantage function influences the overall advantage function. Then, we present the update rule of $\epsilon_k$ during the training process, $\Delta \epsilon_k = (\frac{1}{L} \sum_{l=1}^{L} c_k(l) - \beta) \times (1- \mu) $, in which $\beta$ represents the safety boundary that the agent is penalized for during the training process and $\mu$ is a gain factor. The update formula for $\epsilon_k$ is given by 
\begin{equation}
	\epsilon_k = \text{ReLU}(\epsilon_k + \alpha \Delta \epsilon_k),
\end{equation}
where $\alpha$ is the update coefficient for $\epsilon_k$.
The critic network produces the state value estimate $\hat{V}_{\omega_k}(s(n))$, and it is updated by minimizing the subsequent loss function
\begin{equation}
	J(\omega_k) = \frac{1}{2} \left[ \hat{V}_{\omega_k}(s(n)) - V_k(s(n)) \right]^2.
\end{equation}
Similarly, the cost network can be updated by minimizing the following loss function
\begin{equation}
	J(\varphi_k) = \frac{1}{2} \left[ \hat{C}_{\varphi_k}(s(n)) - C_k(s(n)) \right]^2.
\end{equation}
The MAPPO-Lagrangian algorithm limits policy updates in the actor network by introducing a clipping coefficient $\varepsilon$. Accordingly, the actor loss is formulated as
\begin{equation}
	\begin{aligned}
		J(\theta_k) = & \mathbb{E} \biggl\{ \min \biggl[ \text{clip} \biggl( \frac{\pi_{\theta_k}(a_k(n) \mid s(n))}{\pi_{\theta_k'}(a_k(n) \mid s(n))}, 1-\varepsilon, 1+\varepsilon \biggr) \\
		& \hat{M}_k(s(n)),\frac{\pi_{\theta_k}(a_k(n) \mid s(n))}{\pi_{\theta_k'}(a_k(n) \mid s(n))} \hat{M}_k(s(n)) \biggr] \biggr\},
	\end{aligned}
\end{equation}
where $\theta_k'$ denotes the parameters of the old policy, and $[\pi_{\theta_k}(a_k(n) \mid s(n))$/$ \pi_{\theta_k'}(a_k(n) \mid s(n))]$ is the policy update ratio. Thus, the actor, critic and cost networks can be updated using the gradients $\nabla \theta_k = (\partial J(\theta_k) / \partial \theta_k)$, $\nabla \omega_k = (\partial J(\omega_k) / \partial \omega_k)$ and $\nabla \varphi_k = (\partial J(\varphi_k) / \partial \varphi_k)$, respectively. The above summarizes the MAPPO-Lagrangian training framework, whose pseudocode is shown in Algorithm \ref{alg:mappo-larg}.

\begin{algorithm}[!t]
	\caption{MAPPO-Lagrangian Training Framework}
	\label{alg:mappo-larg}
	\DontPrintSemicolon  
	\SetAlgoLined    
	
	Initialize batch size $B_{\text{batch}}$, number of agents $K$, maximum training episodes $E$, steps per episode $L$, discount factor $\gamma$, PPO update epochs $e_{\text{ppo}}$.\;
	Initialize actor networks $\theta_k$, reward value network $\omega_k$, cost value networks $\varphi_k$, replay buffer $\mathcal{B}_k$, $\forall k \in \mathcal{K}$.\;
	Initialize Lagrangian multipliers $\{\epsilon_k\}_{\forall k \in \mathcal{K}} \leftarrow 0$.\;
	\For{$\text{episode}=1$ \KwTo $E$}{
		\For{$step=1$ \KwTo $\text{L}$}{
			\For{$k=1$ \KwTo $\text{K}$}{
				Obtain observations $o_k$ and global state $s$ from the environment;\;
				Execute actions $a_k$;\;
				Obtain reward $r_k$ from the environment;\;
				The critic network calculates the predicted reward $\hat{V}_{\omega_k}$ based on the global state $s$;\;
				The cost network calculates the predicted penalty $\hat{C}_{\varphi_k}$ based on the global state $s$;\;
			}
			
		}
		\For{$k=1$ \KwTo $\text{K}$}{
			Compute reward advantage function $\hat{A}_k$ and cost advantage function $\hat{A}_k^C$ with generalized advantage estimation;\;
			Sample a random minibatch of $B_k$ transitions from $\mathcal{B}_k$;\;
			\For{$e=1$ \KwTo $e_{\text{ppo}}$}{
				Compute advantage function $\hat{M}_k$;\;
				Adjust $\theta_k$, $\omega_k$ and $\varphi_k$;\;
				Update permanent Lagrangian multipliers $\{\epsilon_k\}$;\;
			}
		}
		
	}
\end{algorithm}

\subsection{Small-Timescale Optimization}
To address the problem $\mathbf{P2}$, we design an improved Discrete Harris Hawks Optimization (DHHO) algorithm. This algorithm eliminates the physically meaningless continuous numerical operations in the traditional HHO algorithm, which adapts to the characteristics of discrete combinatorial optimization, and solves the optimal unloading allocation scheme from the UE to the UAV with the goal of minimizing the total latency of the user task.

The algorithm differentiates the search stages by nonlinearly decaying the energy coefficient $E = 2E_0 \cdot \left(1 - \left(\frac{t}{\mathrm{max\_iter}}\right)^2\right)$, in which $E_0$ is the $[-1,1]$ random number, $t$ is the current iteration number. The nonlinear decay term ensures the convergence of the energy factor in the later iteration, and uses discrete crossover and mutation to achieve optimization. When $|E| \ge 1$, the algorithm enters the global exploration phase. The algorithm selects the update method using a random number $q$, when $q\ge0.5$, it randomly selects other individuals in the population and replaces part of the allocation scheme with a 50\% probability; when $q < 0.5$, it re-randomly allocates available UAVs to some users with a 50\% probability. When $|E| < 1$, it enters the local search phase, and combines the random coefficient $r$ into soft boundary, hard boundary and two diving strategies. It relies on Levy flight to dynamically adjust the mutation probability to optimize the solution, where the standard Levy flight generates continuous random steps. In order to adapt to the UE-UAV discrete allocation optimization, DHHO gives up the continuous position update and maps the Levy random step value to the gene mutation probability. If the population individuals coincide with the optimal solution, random mutation is performed to prevent premature update, and the optimal assignment result is continuously updated iteratively. After that, the illegal allocation was corrected and the faulty UAV number was replaced. Then, the users with the highest delay are redistributed and optimized by heuristic repair, and the fitness is the total delay plus timeout penalty.

When the DHHO algorithm is iterated, the UAV computing resource allocation is optimized with the goal of minimizing the delay, and then the matching strategy between the UE and the UAV is quantitatively evaluated. The scenario is that the UAV has identified the UEs to serve and will decide how to allocate its computing resources to minimize the total task processing latency, as 
\begin{subequations}
\begin{align}
&\min_{f_{m,u}^t} \sum_{m \in \mathcal{M}_0} \frac{W_m(t)}{f_{m,u}^t}\\
\text{s.t.} ~ &\sum_{m \in \mathcal{M}_0} f_{m,u}^t = \varpi_u, ~ \forall u \in \mathcal{U}, \label{eq:f}\\
&f_{m,u}^t \geq 0, ~ \forall m \in \mathcal{M}_0, \forall u \in \mathcal{U}, \label{eq:f0}
\end{align}
\end{subequations}
where $\mathcal{M}_0$ represents the set of UEs whose tasks are uploaded for processing by the UAV and $W_m(t) = D_m(t) C_m(t)$ denotes the total computational workload of UE $m$ task in time slot $t$.

The second derivative of the $\frac{W_m(t)}{f_{m,u}^t}$ with respect to $f_{m,u}^t$ is $\frac{2W_m(t)}{(f_{m,u}^t)^3} > 0$, which is a convex function. The sum of convex functions remains convex, so the objective function is convex.
The constraint (\ref{eq:f}) is an affine equality constraint, and (\ref{eq:f0}) is a convex constraint.
Therefore, the problem is convex, ensuring that any locally optimal solution is globally optimal.
We solve the computing resource allocation subproblem using the Karush-Kuhn-Tucker (KKT) method, since it effectively handles complexity by reformulating the original problem into an optimization problem with equality constraints, thereby improving solution efficiency when multiple constraints are involved.

We construct the Lagrangian function by introducing the Lagrange multiplier $\lambda \geq 0$, as
\begin{equation}
\begin{aligned}
\mathcal{L}(f_{1,u}^t, \dots, f_{M,u}^t, \lambda) &= \\
&\sum_{m \in \mathcal{M}_0} \frac{W_m(t)}{f_{m,u}^t} + \lambda \left( \sum_{m \in \mathcal{M}_0} f_{m,u}^t - \varpi_u \right)
\end{aligned}
\end{equation}
The constraint (\ref{eq:f0}) can be ignored because the optimal solution satisfies (\ref{eq:f0}), otherwise the latency would be infinite, and thus the corresponding multiplier is zero under the complementary slackness condition.

We take the partial derivative with respect to $f_{m,u}^t$ and set it to zero to obtain the extremum condition, i.e.,
\begin{equation}
\frac{\partial \mathcal{L}}{\partial f_{m,u}^t} = -\frac{W_m(t)}{(f_{m,u}^t)^2} + \lambda = 0. \label{39}
\end{equation}
Based on (\ref{39}), it can be concluded that
\begin{equation}
\lambda = \frac{W_m(t)}{(f_{m,u}^t)^2} \implies f_{m,u}^t = \sqrt{\frac{W_m(t)}{\lambda}}.
\end{equation}
Substituting the result into the constraint (\ref{eq:f0}), we get
\begin{equation}
	f_{m,u}^t = \frac{\sqrt{W_m(t)} \cdot \varpi_u}{S},
\end{equation}
where $S = \sum_{m \in \mathcal{M}_0} \sqrt{W_m(t)}$. This is the core closed-form solution, the optimal resource allocation for UE $m$ is proportional to $\sqrt{W_m(t)}$ and inversely proportional to the sum of $\sqrt{W_m(t)}$ for all UEs. The above summarizes the training framework of DHHO combined with convex optimization, whose pseudocode is shown in Algorithm \ref{alg:dhho}.

\begin{algorithm}[!t]
	\caption{DHHO Algorithm for UAV Task Allocation}
	\label{alg:dhho}
	\DontPrintSemicolon
	\SetAlgoLined
	
	Initialize population size $N$, maximum iteration $T$, active UAV set $\mathcal U$.\;
	Precompute transmission rate, task parameters $W,\sqrt{W}$.\;
	Generate initial greedy solution $\boldsymbol X_{\text{greedy}}$.\;
	Initialize population $\{\boldsymbol X_i\}_{i=1}^N$ based on greedy solution and random initialization.\;
	Set rabbit solution $\boldsymbol X^*=\boldsymbol X_{\text{greedy}}$, optimal fitness $F^*=\infty$, convergence list $\mathcal C=\emptyset$.\;
	
	\For{$t=1$ \KwTo $T$}{
		\For{$i=1$ \KwTo $N$}{
			Evaluate fitness $F_i$ via objective function with heuristic repair;\;
			\If{$F_i<F^*$}{$F^* \leftarrow F_i,\boldsymbol X^*\leftarrow \boldsymbol X_i$;}
		}
		Add $F^*$ into $\mathcal C$;\;
		Calculate nonlinear decay energy factor $E$;\;
		\For{$i=1$ \KwTo $N$}{
			$\boldsymbol X_{\text{new}}\leftarrow \boldsymbol X_i$;\;
			\If{$|E|\ge1$}{
				$q \leftarrow \mathrm{rand}(0,1)$;\;
				\If{$q \ge 0.5$}{
					Pick $\boldsymbol X_k$ randomly;\;
					$\boldsymbol m \leftarrow \mathbb{I}\{\mathrm{rand}<0.5\}$;\;
					$\boldsymbol X_{\text{new}}(\boldsymbol m) \leftarrow \boldsymbol X_k(\boldsymbol m)$;\;
				}
				\Else{
					$\boldsymbol m \leftarrow \mathbb{I}\{\mathrm{rand}<0.5\}$;\;
					$\boldsymbol X_{\text{new}}(\boldsymbol m) \leftarrow \mathrm{rand}(\mathcal U)$;\;
				}
			}\Else{
				\If{$r\ge0.5$}{Soft/hard surround update toward $\boldsymbol X^*$;}
				\Else{Tentative update + Levy mutation for rapid dive exploitation;}
			}
			$\boldsymbol X_i\leftarrow \boldsymbol X_{\text{new}}$;\;
		}
	}
	\Return optimal allocation $\boldsymbol X^*$, convergence curve $\mathcal C$.\;
\end{algorithm}

\subsection{Computational Complexity Analysis}
The computational complexity of the proposed two-timescale optimization framework is analyzed by sequentially evaluating the large-timescale trajectory control algorithm and the small-timescale task scheduling algorithm. For the large-timescale multi-agent proximal policy optimization-Lagrangian framework, the computational overhead is primarily dominated by the forward and backward propagation processes of the deep neural networks. Denoting the maximum training episodes as $E$, the steps per episode as $L$, the number of UAVs as $U$, the proximal policy optimization update epochs as $e_{\text{ppo}}$, the batch size as $B$, the number of fully connected layers as $L_\text{nn}$, and the maximum number of neurons per layer as $N_\text{nn}$, the overall centralized training time complexity is strictly bounded by $\mathcal{O}(E L U e_\text{ppo} B L_\text{nn} N_\text{nn}^2)$. During the decentralized execution phase, each uncrewed aerial vehicle agent solely relies on the forward pass of its local actor network to generate real-time trajectory decisions, yielding a highly lightweight execution complexity of $\mathcal{O}(L_\text{nn} N_\text{nn}^2)$ per large time slot. In parallel, the small-timescale optimization manages task reassignment and resource allocation utilizing the improved Discrete Harris Hawks Optimization algorithm combined with convex optimization. Defining the heuristic population size as $N$, the maximum number of iterations as $T$, and the total number of UE as $M$, the algorithm evaluates the fitness of each discrete individual by calculating the closed-form Karush-Kuhn-Tucker solution for computing resource allocation. Since this closed-form evaluation computes the optimal frequency allocation inversely proportional to the workload sum, it incurs a linear complexity proportional to the product of $M$ and $U$ across the network. Consequently, the total time complexity for the small-timescale task scheduling process is securely bounded by $\mathcal{O}(T N M U)$. By physically decoupling the highly complex neural network training from the online execution phase and leveraging a low-complexity linear closed-form solution within the heuristic search loop, the proposed overall algorithmic architecture achieves a scalable and strictly polynomial execution time complexity.

\section{Numerical Results}\label{Sec.Results}
This section provides simulation experiments to demonstrate the effectiveness of our proposed algorithm in multi-UAV assisted MEC networks within a two-timescale training framework. In the simulation, the UAV service area is configured as a 500 m × 500 m square region with an altitude constraint of [50, 80] m. The initial positions of the UAVs are randomly placed within the service volume, subject to the minimum safe separation distance between UAVs. All random realizations of UE and UAV deployments are independently drawn for each simulation trial. The probability that each UAV is unavailable at each small time slot $t$ is 0.5\%. All networks in MAPPO-Lagrangian use a lightweight fully connected multi-layer perceptron architecture, with two hidden layers of 128 neurons each and ReLU activation functions. For the actor network that outputs continuous velocity commands, the output layer generates the mean of each action dimension, paired with learnable logarithmic standard deviations to form the stochastic policy. The reward critic and cost critic networks have the same hidden structure and output scalar estimates of state value and state cost, respectively. The PPO clipping threshold is set to 0.2, and policy updates are performed for 10 epochs per collected batch to improve sample efficiency. For the Lagrangian multipliers that enforce constraints, the safety boundary $beta$, gain factor $\mu$, and update coefficient $alpha$ are set to 0.0, 0.01, and 0.05, respectively, tuned to gradually tighten constraint satisfaction without disrupting policy improvement. To stabilize training and resolve the scale mismatch between latency and energy terms in the reward function, we apply running mean-std normalization to all reward signals. The maximum number of training rounds is 500 rounds, the length of each round is 128 steps, the discount factor is 0.95, the reward discount factor is 0.99, and the optimizer we use is Adam.
The additional simulation parameters are presented in Table \ref{tab:sim_param}, in line with prior work.
\begin{table}[t!]
	\caption{\normalsize Simulation Parameters}
	\label{tab:sim_param}
	\centering
	\renewcommand{\arraystretch}{1.05}
	\footnotesize
	\begin{tabularx}{\columnwidth}{@{}c>{\raggedright\arraybackslash}Xc@{}}
		\toprule
		\textbf{Symbol} & \textbf{Description} & \textbf{Value} \\
		\midrule
		$\mathcal{M}$ & Set of UEs & $20$ \\
		$\mathcal{U}$ & Set of UAVs & $5$ \\
		$D_{\text{min}}$ & Safe distance between UAVs & $3~\text{m}$ \\
		$v_{\text{max}}$ & Maximum UAV speed & $50~\text{m/s}$ \\
		$P_0$ & Blade profile power & $80~\text{W}$ \\
		$P_{\text{ind}}$ & Induced power & $100~\text{W}$ \\
		$U_{\text{tip}}$ & Rotor blade tip speed & $120~\text{m/s}$ \\
		$v_0$ & Average induced velocity in hover & $4~\text{m/s}$ \\
		$d_0$ & Fuselage drag coefficient & $0.6$ \\
		$\rho_0$ & Air density & $1.225~\text{kg}/\text{m}^3$ \\
		$s$ & Rotor solidity & $0.08$ \\
		$A$ & Area swept by rotor blades & $0.5~\text{m}^2$ \\
		$W$ & Weight of the UAV & $2~\text{kg}$ \\
		$g$ & Gravitational acceleration & $9.8~\text{m}/\text{s}^2$ \\
		$f_a$ & A2A carrier frequency & $5.8~\text{GHz}$ \\
		$f_c$ & A2G carrier frequency & $2.4~\text{GHz}$ \\
		$\varpi_u$ & Max computing capacity of UAV $u$ & $5~\text{GHz}$ \\
		$N_0$ & Noise power density & $3.98\times10^{-21}~\text{W/Hz}$ \\
		$B$ & System bandwidth & $10~\text{MHz}$ \\
		$\eta_{\text{LoS}}$ & Excess path loss (LoS) & $1.0~\text{dB}$ \\
		$\eta_{\text{NLoS}}$ & Excess path loss (NLoS) & $20~\text{dB}$ \\
		\bottomrule
	\end{tabularx}
\end{table}

The performance of the proposed algorithm is compared with the following baselines.
\begin{enumerate}
	\item \textit{Particle swarm optimization (PSO):} This algorithm simulates the foraging behavior of bird flocks \cite{chen2022joint}. It uses a group of particles representing potential solutions, which dynamically adjust based on their own historical best positions and the global best positions of the group. It balances local exploration and global exploitation, collaboratively searching for the optimal combination of UAV trajectories, task allocation, and resource allocation, in order to maximize the total reward of the system.
	\item \textit{Greedy:} This algorithm, based on the current knowledge, greedily selects the trajectory of the UAV, the task division, and the allocation of computing resources, in order to minimize the system loss to the greatest extent \cite{xu2022cellular}.
	\item \textit{MAPPO-Greedy:} This algorithm is a combination of the greedy algorithm and the MAPPO algorithm \cite{guan2023mappo}. It uses the greedy algorithm to make decisions on task division and the allocation of computing resources for UAVs to achieve the maximum minimization of latency. Additionally, it combines the trajectory optimization of the MAPPO algorithm to explore the optimal situations at different locations, in order to maximize the total reward of the system.
	\item \textit{MAPPO-Lagrangian-GWO:} This algorithm is a combination of GWO and MAPPO-Lagrangian algorithms \cite{liu2024heterogeneous}. The GWO algorithm simulates the collaborative hunting behavior of wolves in a social hierarchy. Through the guidance of three elite wolves, alpha, beta, and delta, the entire wolf pack searches in the solution space. It combines high-quality greedy seed initialization and two-stage heuristic local repair strategies to iteratively optimize the user task allocation scheme, in order to minimize the total system delay and timeout penalty to the greatest extent, and combines the MAPPO-Lagrangian algorithm to optimize the UAV trajectory.
	\item \textit{Multi-agent deep deterministic policy gradient-grey wolf optimizer (MADDPG-GWO):} This algorithm is a combination of GWO and MADDPG \cite{zhao2022multi}. Among them, MADDPG is made up of two actor and two critic networks, and the action value derived from the actor networks is subsequently augmented by exploration noise. Subsequently, the critic networks evaluate the action-value function and use it to optimize the trajectory of the UAV.
	\item \textit{Hierarchical MAPPO:} Among them, MAPPO is employed separately to optimize the optimization variables of the large and small time slots. Specifically, the hierarchical MAPPO consists of a high-level MAPPO and a low-level MAPPO, which respectively optimize the optimization variables in the large and small time slots.
	
\end{enumerate}

\begin{figure}[t!]
	\centering
	\includegraphics[width=0.5\textwidth]{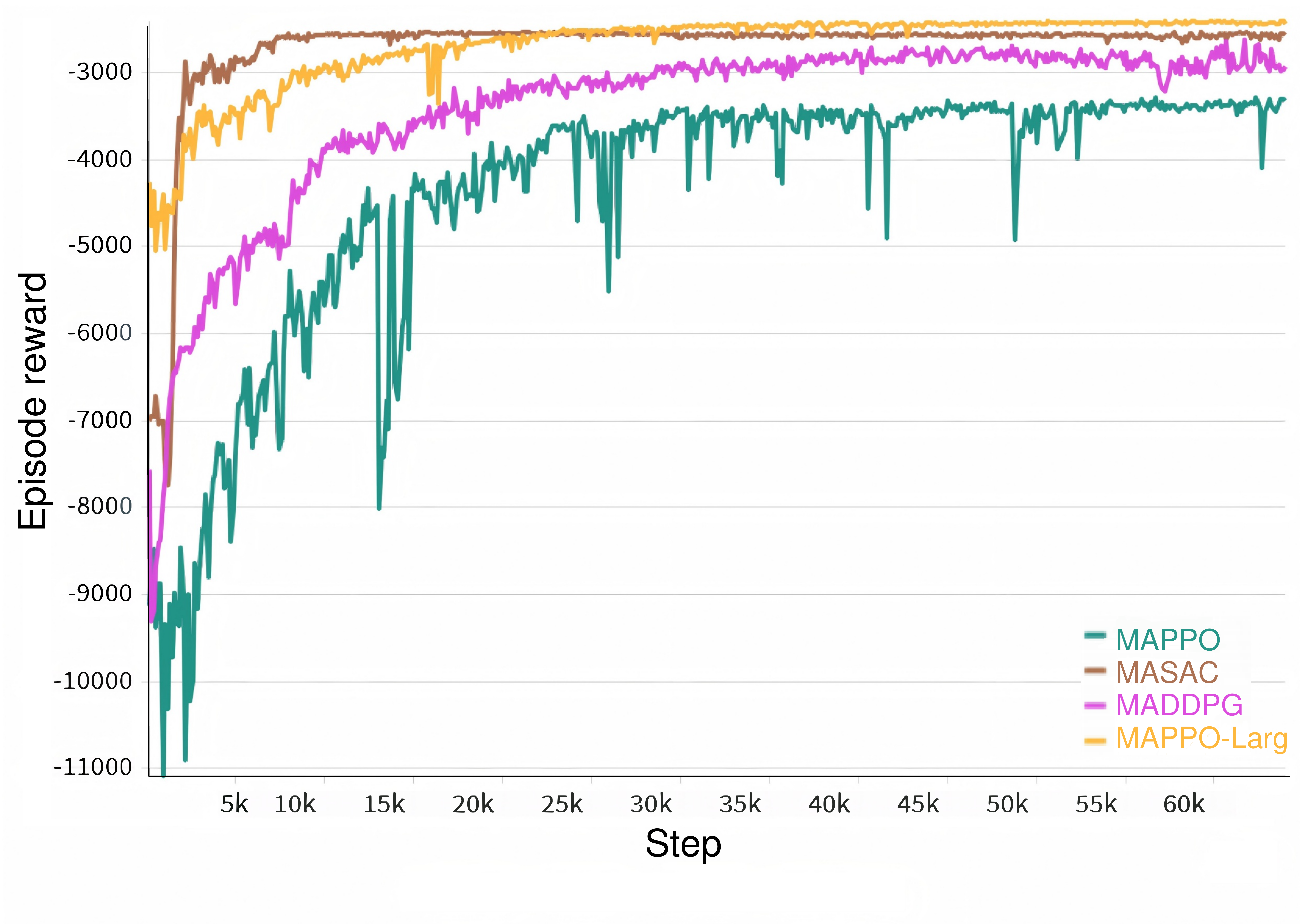}
	\caption{Convergence situations of different algorithms in large time slots.}
	\label{fig. p1}
	\vspace{0 cm}
\end{figure}

\begin{figure}[t!]
	\centering
	\includegraphics[width=0.5\textwidth]{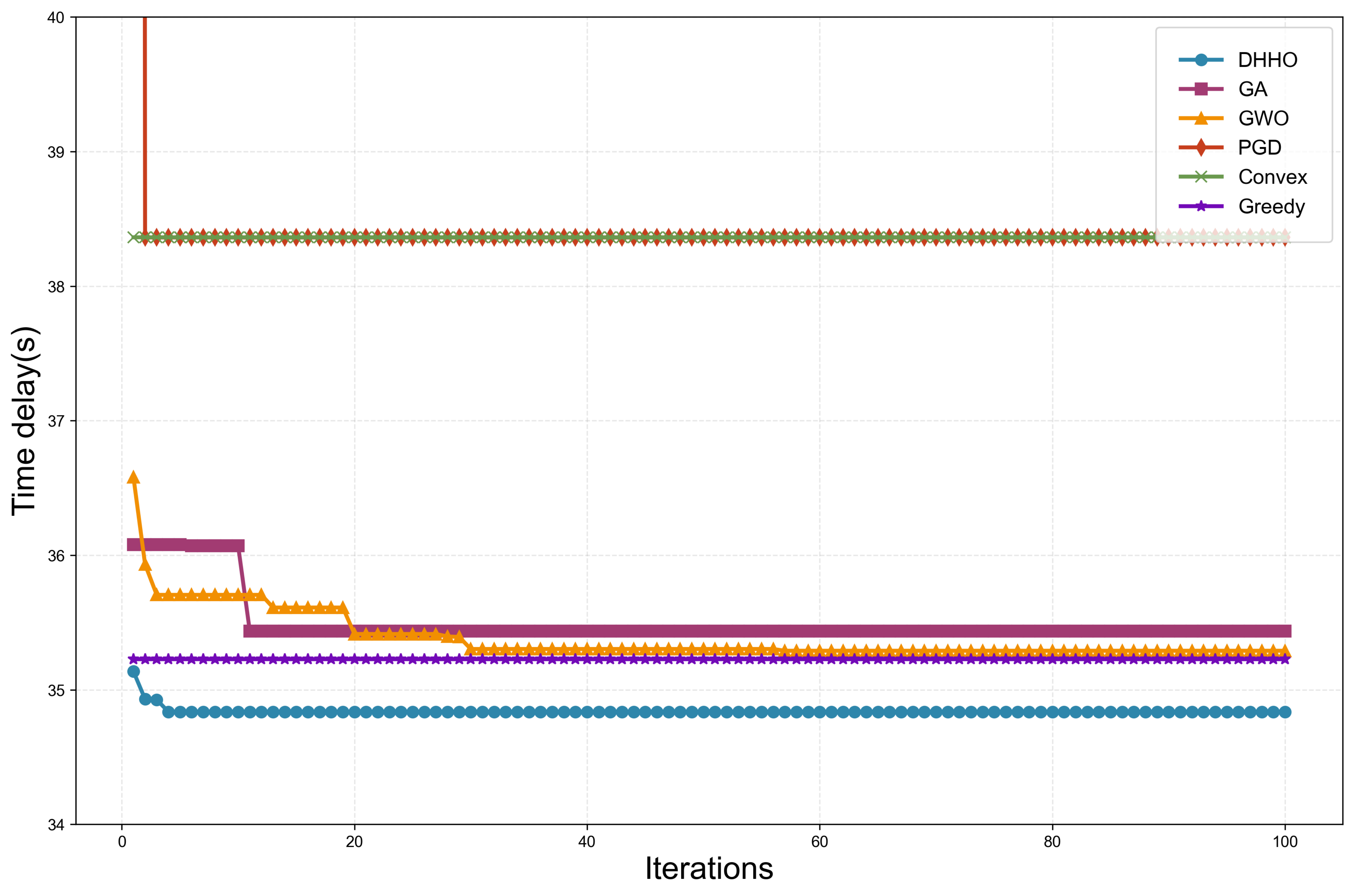}
	\caption{The delay situations obtained through different algorithms' iterations in small time intervals.}
	\label{fig. p2}
	\vspace{0 cm}
\end{figure}
In Fig. \ref{fig. p1}, we report the convergence of our MAPPO-Lagrangian algorithm in relation to other benchmarks. Meanwhile, all algorithms achieve progressively higher rewards with increasing training iterations, confirming that MADRL is effective for offloading computations. Moreover, it is evident that the MAPPO-Lagrangian algorithm outperforms the original MAPPO and MADDPG algorithms by achieving the highest rewards and exhibiting a faster convergence speed. This demonstrates that in our network, the additional Lagrange multipliers and loss network we added can further enhance the performance of the MAPPO algorithm. Unlike the conventional MAPPO that incorporates constraints merely as penalty terms within the reward function leading to conflicting gradients and unstable exploration, the MAPPO-Lagrangian framework explicitly models the system as a CMDP. By introducing independent cost critic networks and adaptive Lagrangian multipliers for each agent, our method decouples the optimization of system utility from constraint satisfaction. This not only stabilizes the training process via more accurate advantage estimation but also enables fine-grained credit assignment and robust handling of UAV failures and position offsets. The adaptive adjustment of $\epsilon_k$ effectively prevents overly conservative or overly aggressive policies, which is why our approach consistently outperforms vanilla MAPPO in both latency reduction and cost saving. Fig. \ref{fig. p2} shows the changes in delay loss for different algorithms during 100 iterations. Our proposed improved DHHO algorithm achieves very good results in the early stages of the iterations, indicating that incorporating the results of the greedy algorithm as the initial seed can effectively improve the performance of the algorithm.

\begin{figure}[t!]
	\centering
	\includegraphics[width=0.5\textwidth]{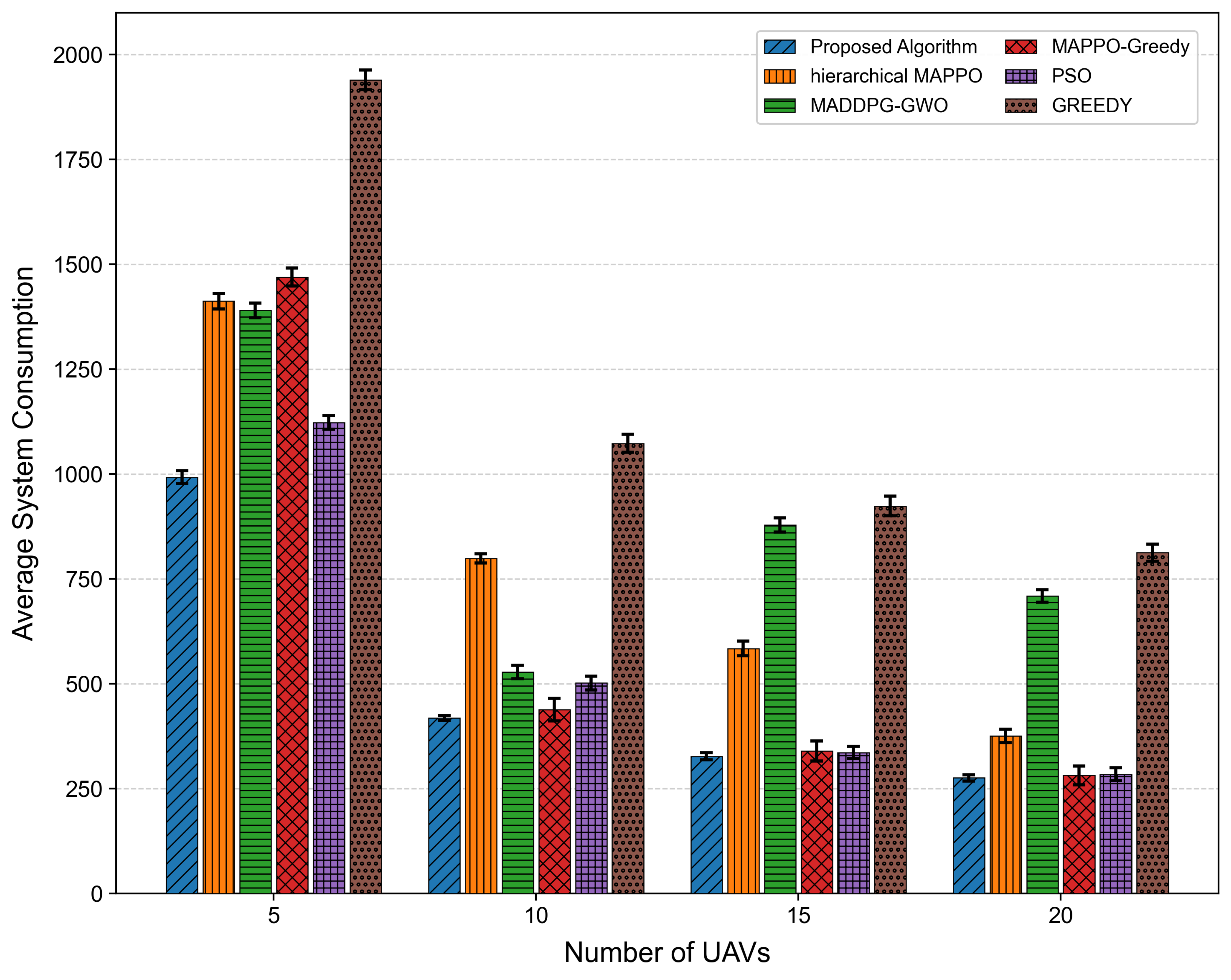}
	\caption{Compare the average system losses obtained by different algorithms under different numbers of UAVs. (10 independent runs; 95\% confidence intervals)}
	\label{fig. p3}
	\vspace{0 cm}
\end{figure}

Fig. \ref{fig. p3} compares the average system loss of six algorithms under different numbers of UAVs when there are 20 UEs. The weighted average system loss shows a notable downward trend when the UAV count increases. This phenomenon can be explained as follows: increasing the number of UAVs leads to a corresponding expansion of the available computational capacity. This enables the agents to better distribute user tasks more reasonably, thereby achieving a lower weighted average system loss. In addition, our solution demonstrates consistent superiority over the other five solutions across all settings, and this advantage is even more significant when UAV resources are scarce.

\begin{figure}[t!]
	\centering
	\includegraphics[width=0.5\textwidth]{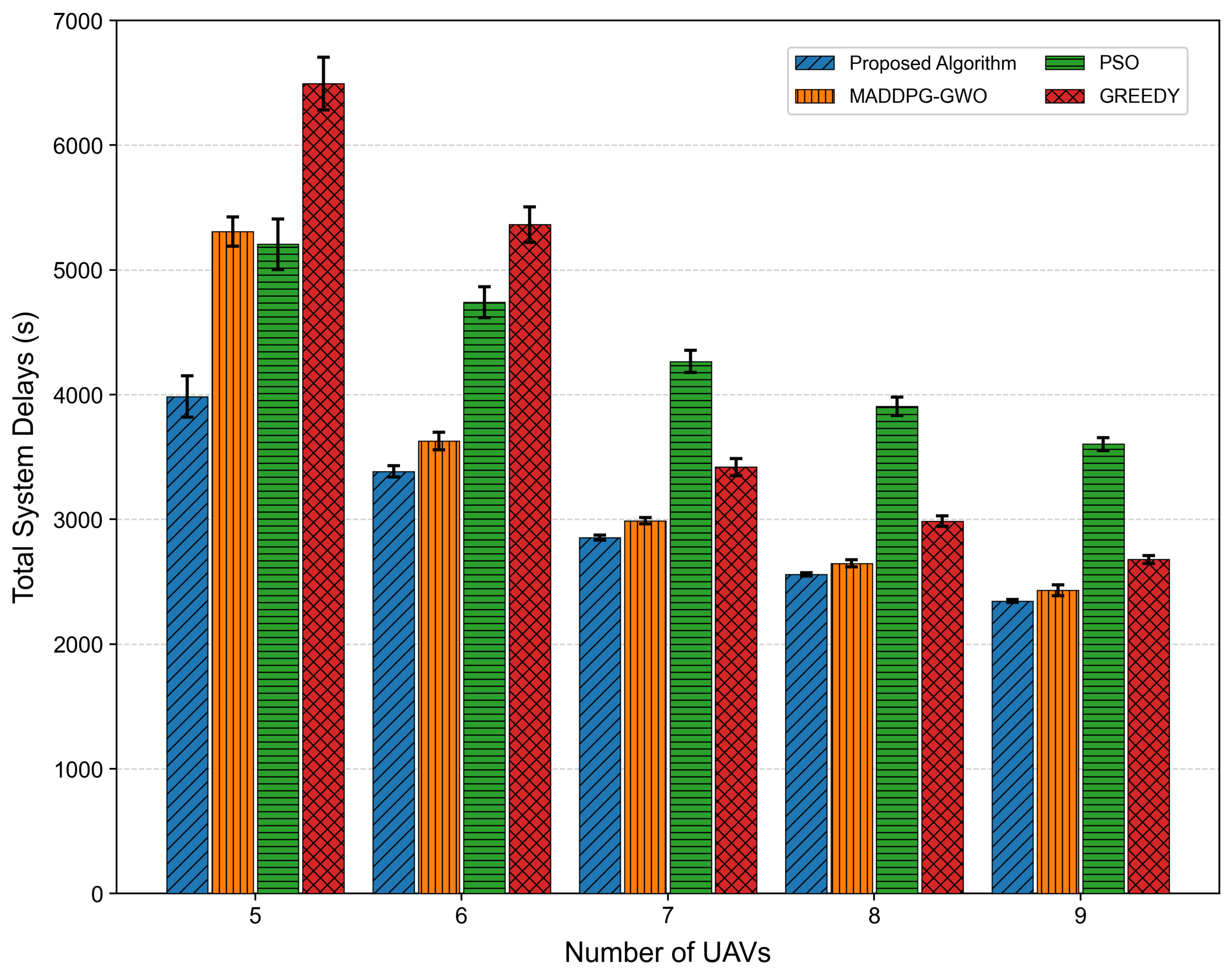}
	\caption{Compare the total system latency obtained by different algorithms under different numbers of UAVs.}
	\label{fig. p4}
	\vspace{0 cm}
\end{figure}

\begin{figure}[t!]
\centering
\includegraphics[width=0.5\textwidth]{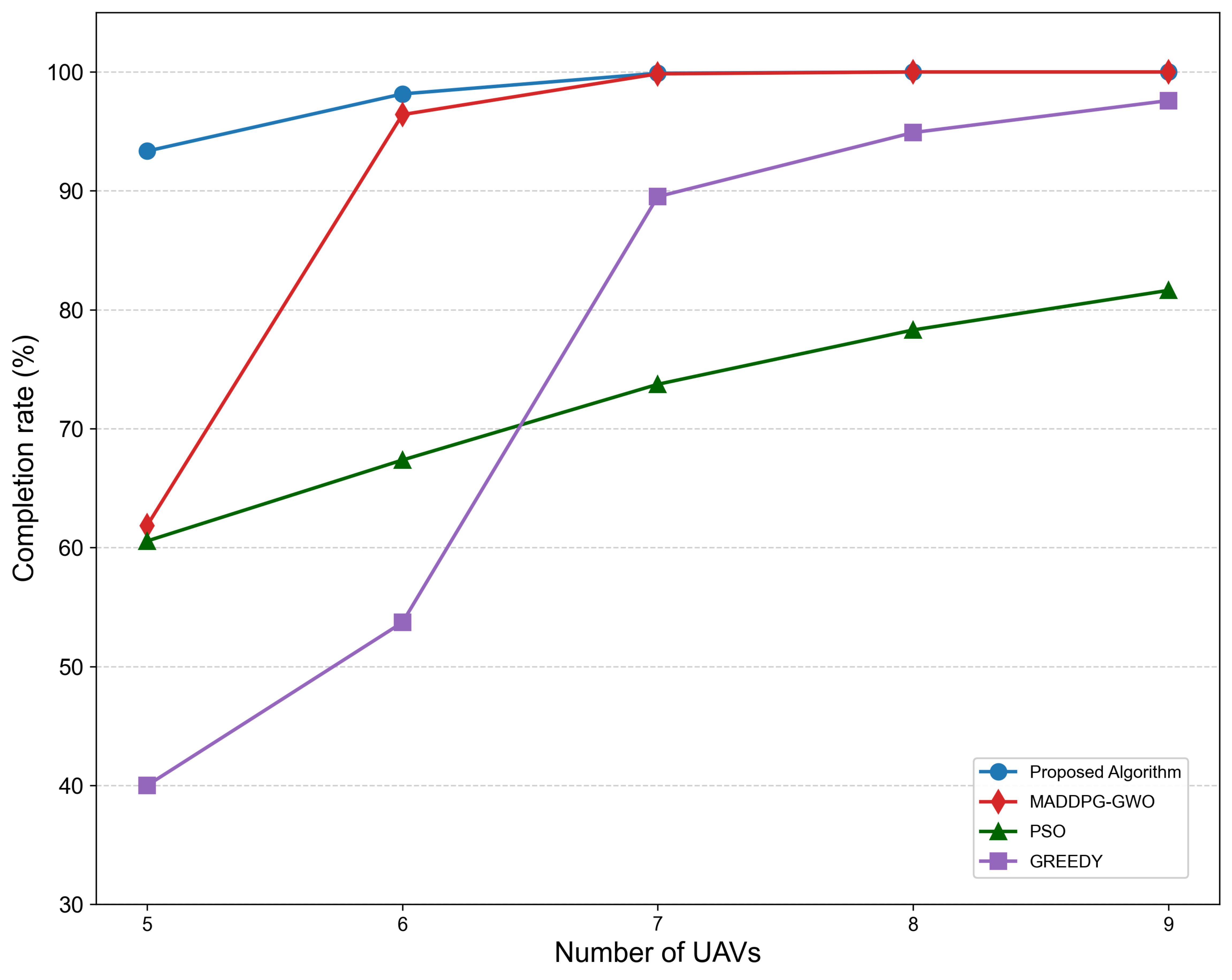}
\caption{Compare the task completion rates of different algorithms under different numbers of UAVs.}
\label{fig. p5}
\vspace{0 cm}
\end{figure}
Fig. \ref{fig. p4} compares the total system delay of the six algorithms with different numbers of UAVs in the case of 20 UEs. As the number of UAVs increases, the overall system delay shows a downward trend. Due to the increase of the number of UAVs, the agent can allocate UAVs and computing resources to tasks more reasonably, so as to obtain a smaller overall system delay. It can be noticed that our algorithm obtains the smallest overall system delay in all cases. Combined with Fig. \ref{fig. p5}, we can see that with the increase of the number of UAVs, the completion rate of the task also increases step by step. And it can be noticed that when the number of UAVs reaches 7 and beyond, although the task completion rate of our algorithm and MADDPG-GWO both reach 100\%, our algorithm is still ahead of MADDPG-GWO in the overall system delay. And when the number of UAVs is small, it has more obvious advantages than other MADDPG-GWO or Greedy, which reflects that our DHHO algorithm can still make better decisions on task offloading and allocation of computing resources in the case of tight computing resources, and can make better decisions than other algorithms when the computing resources are sufficient. Compared with the traditional scheme of ``continuous variable optimization + integer mapping'', in order to adapt to the discrete variable problem of task allocation, our DHHO algorithm directly searches the optimization variable in the integer solution space, completely eliminates the invalid search space, and the operation has clear physical meaning.
\begin{figure}[t!]
	\centering
	\includegraphics[width=0.5\textwidth]{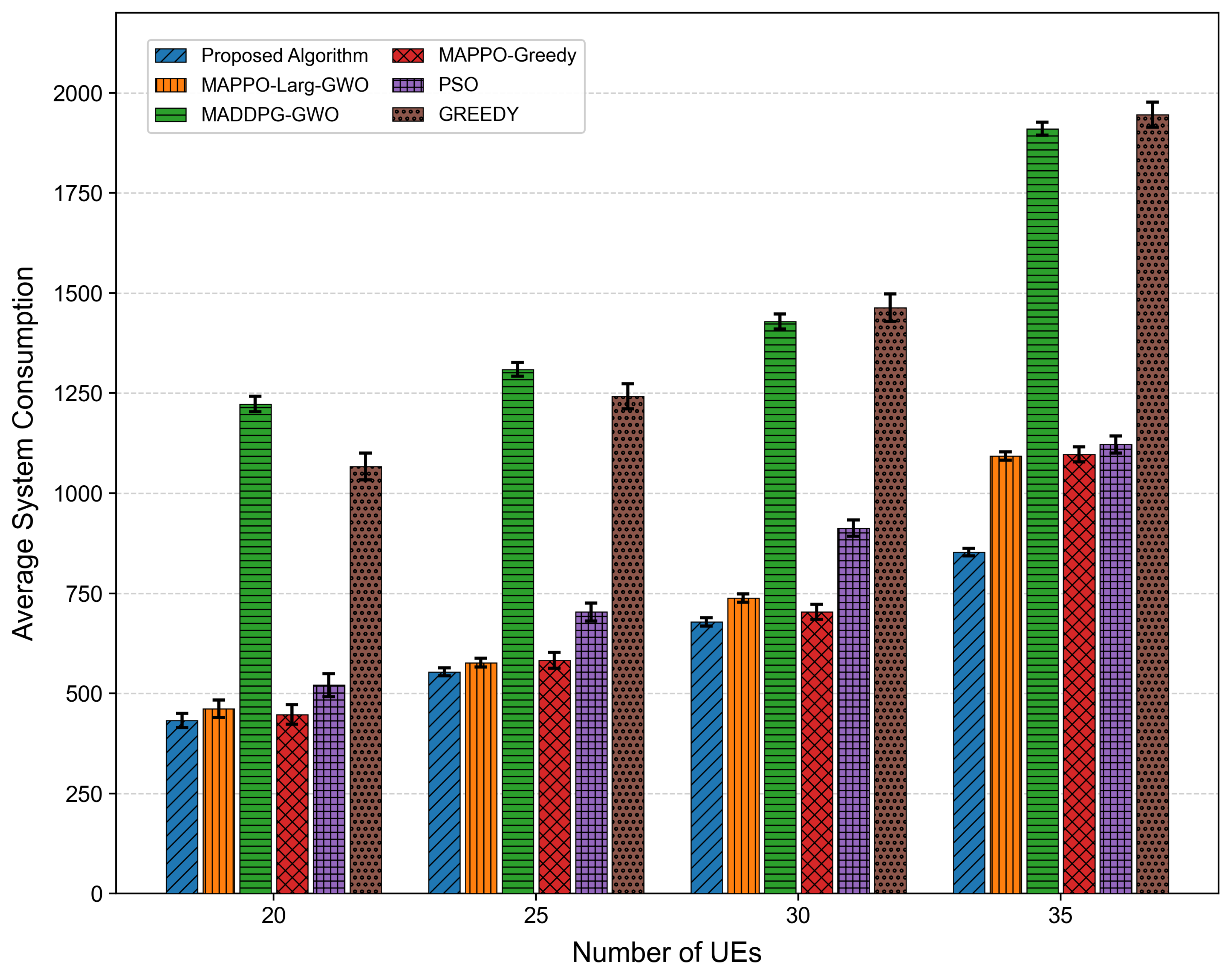}
	\caption{Compare the average system loss of different algorithms with different number of UEs. (10 independent runs; 95\% confidence intervals)}
	\label{fig. p6}
	\vspace{0 cm}
\end{figure}
Fig. \ref{fig. p6} compares the average system loss under different numbers of UEs. According to the results, the DRL algorithm exhibits better performance than the greedy algorithm. The reason is that DRL is capable of adapting to an uncertain environment via continuous interaction, whereas the greedy algorithm is prone to getting stuck in suboptimal local solutions. Moreover, the performance of our algorithm is superior to PSO and MAPPO-Greedy algorithms, and there is a significant performance gap between our algorithm and MADDPG-GWO algorithm. Additionally, as the number of UEs increases, the average system loss also increases. This is because more UEs require more computing and communication resources, and serving different UEs may cause the UAV to fly more frequently in different areas, thereby increasing energy consumption.

\begin{figure}[t!]
	\centering
	\includegraphics[width=0.5\textwidth]{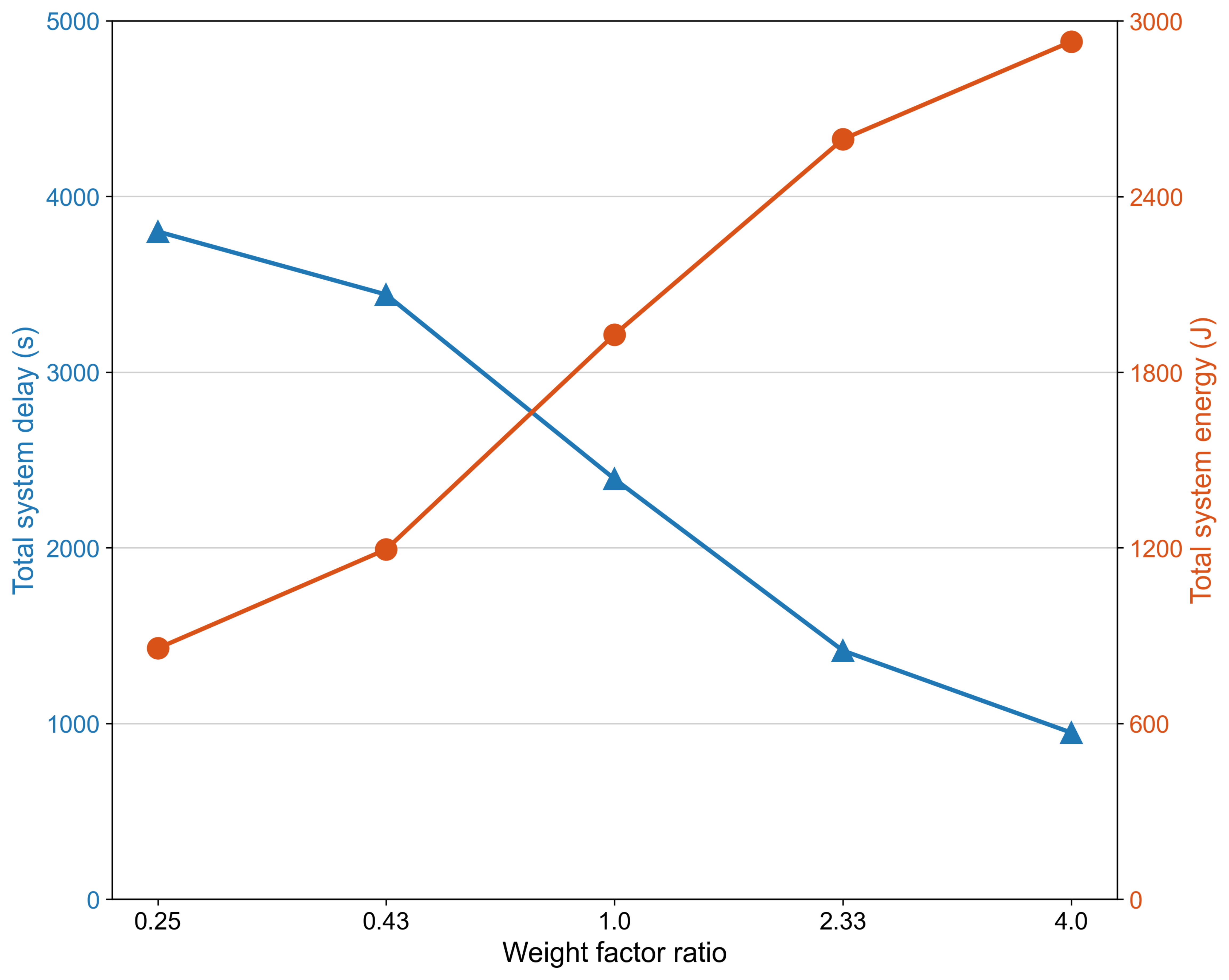}
	\caption{Impact of the weighting factor on delay and energy consumption.}
	\label{fig. w1_w2}
	\vspace{0 cm}
\end{figure}

The ratio of the delay weight $w_1$ to the energy consumption weight $w_2$ determines the priority between the two types of metrics in multi-objective optimization, and its variation exhibits a clear performance trade-off pattern. As can be seen from Fig. \ref{fig. w1_w2}, as the ratio increases from 0.25 to 4.0, the total system delay decreases monotonically, while the total energy consumption continues to rise. At low ratios, energy consumption dominates the optimization. The large-timescale trajectory optimization restricts the UAV's maneuvering amplitude to reduce flight energy consumption, and the small-timescale task scheduling increases the task processing time because the UAV has not reached a reasonable position, which correspondingly leads to higher delay. At high ratios, the optimization favors delay. The UAV actively moves toward user-dense areas to shorten communication distances, while both task reassignment and resource scheduling focus on minimizing delay, significantly improving delay performance but correspondingly increasing flight energy consumption. These results verify the flexibility of the linear weighted multi-objective optimization method.
		
\begin{figure}[t!]
	\centering
	\includegraphics[width=0.5\textwidth]{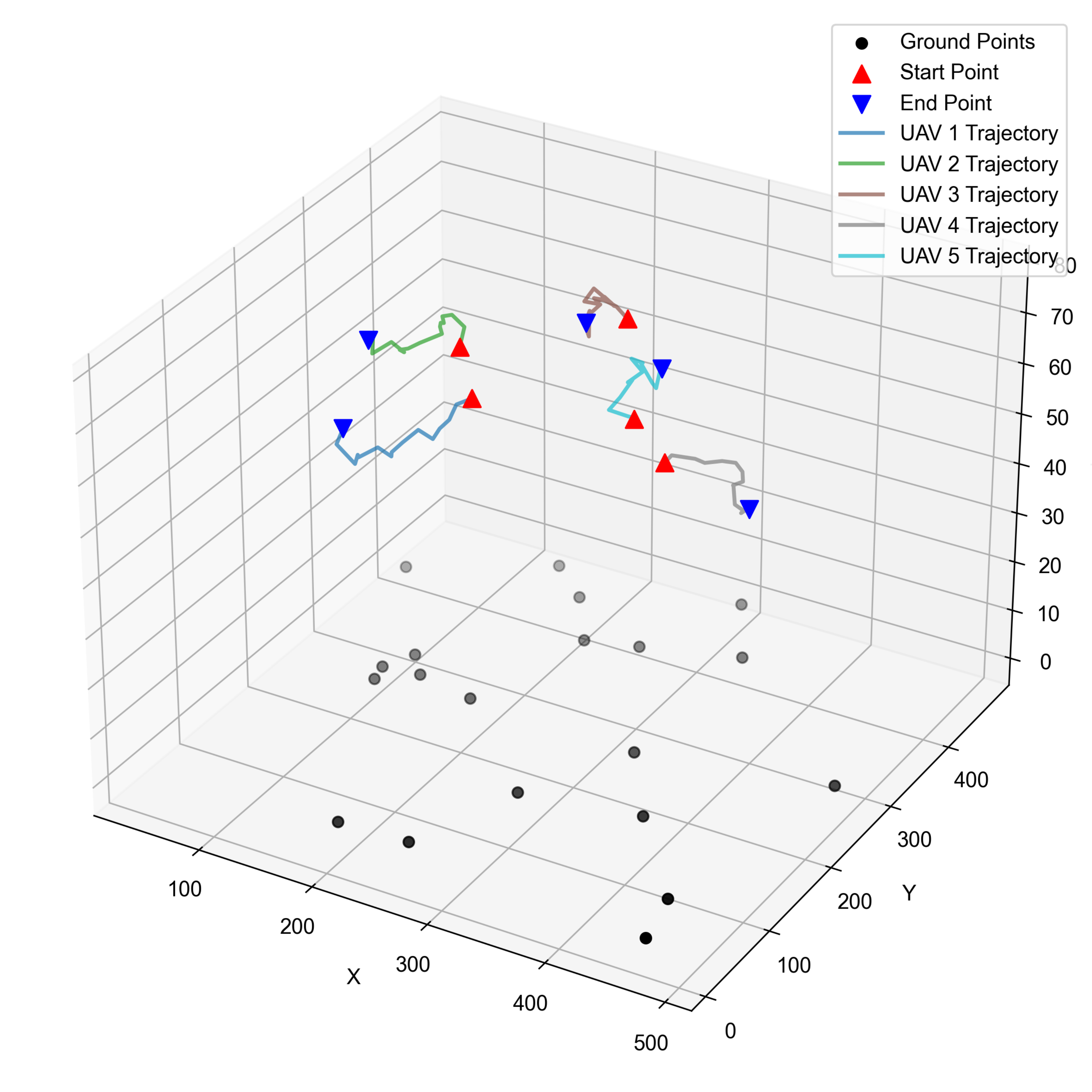}
	\caption{Example of trajectories of UAVs.}
	\label{fig. 3d}
	\vspace{0 cm}
\end{figure}

In Fig. \ref{fig. 3d}, we present the trajectory of the UAV. Clearly, the UAV has the ability to identify areas with high UE concentration and adjust its position accordingly based on the UE distribution. Since the reward mechanism helps the agent determine where the latency is lower, the UAV gradually finds better areas to serve users through rewards.

\begin{table}[t]
	\centering
	\caption{Ablation Results of the Proposed Modules}
	\label{tab:ablation results}
	\begin{tabular}{l r r r}   
		\hline
		\textbf{Module} & \textbf{Reward} & \textbf{Delay(s)} & \textbf{Energy(J)} \\
		\hline
		Full            & 4376.20 & 4783.22 & 3857.16 \\
		No transfer     & 5521.90 & 7074.04 & 3785.76 \\   
		Unrestricted DRL & 4680.50 & 4801.66 & 4559.34 \\
		No heuristics   & 5878.46 & 6711.84 & 4621.06 \\
		No CVX          & 5140.81 & 6,375.30 & 3,786.32 \\
		\hline
	\end{tabular}
\end{table}

The ablation results in Table \ref{tab:ablation results} verify the effectiveness of the key modules. Removing the task reassignment mechanism leads to significant degradation in system delay and overall performance, indicating its critical role in load balancing among UAVs and latency reduction; the slight decrease in energy consumption mainly stems from reduced communication and maneuvering overhead caused by fewer task transfers. Replacing the proposed constrained reinforcement learning framework with unconstrained MAPPO results in a notable increase in energy consumption and makes it difficult to strictly satisfy hard constraints such as safe distance and speed boundaries, suggesting that explicitly modeling the CMDP and incorporating Lagrange multipliers is essential for constraint satisfaction. Removing the heuristic optimization module or the convex optimization module would significantly increase the complexity of decision-making, leading to an increase in processing delay for the task and a decrease in the overall system performance. These results indicate that the task re-allocation mechanism, constraint reinforcement learning, and heuristic convex optimization are all indispensable for improving system performance and ensuring constraint satisfaction.

\section{Conclusion}\label{Sec.Conclusion}
In this paper, we have proposed a robust two-timescale multi-UAV-assisted MEC computing offloading scheme, considering that the task arrival frequency and the UAV action decision frequency are inconsistent. We have constructed a system-level weighted minimization problem of energy consumption and delay by optimizing UAV trajectories in large time slots and optimizing task offloading and computation resource allocation in small time slots. To solve this resultant problem, we have developed a two-timescale training framework, which employs MAPPO-Lagrangian combined with an improved DHHO algorithm to solve the respective problems in the two time dimensions. Extensive numerical results have shown that the proposed scheme outperforms the benchmark methods in reducing energy consumption. In future research, we will explore the task offloading problem of UAVs in the fusion scenario of communication and perception, and explore more practical applications of UAVs, such as the use in mechanical control or video streaming services.

\bibliographystyle{IEEEtran}
\bibliography{transreference}

@article{11021435,
	title={Dynamic Offloading Based on Lyapunov Optimization for UAV-Assisted Maritime IoT-MEC Networks}, 
    author={Bai, Yuting and Zhang, Ying},
	journal={IEEE Transactions on Vehicular Technology}, 
	year={2025},
	volume={74},
	number={11},
	pages={17894-17906},
	publisher={IEEE}
}

@article{li2024uav,
	title={{UAV}-enabled fair offloading for {MEC} networks: A {DRL} approach based on actor-critic parallel architecture},
	author={Li, Wei and Li, Si and Shi, Huaguang and Yan, Wenhao and Zhou, Yi},
	journal={Applied Intelligence},
	volume={54},
	number={4},
	pages={3529--3546},
	year={2024},
	publisher={Springer}
}

@article{wang2021deep,
	title={Deep reinforcement learning based dynamic trajectory control for {UAV}-assisted mobile edge computing},
	author={Wang, Liang and Wang, Kezhi and Pan, Cunhua and Xu, Wei and Aslam, Nauman and Nallanathan, Arumugam},
	journal={IEEE Transactions on Mobile Computing},
	volume={21},
	number={10},
	pages={3536--3550},
	year={2021},
	publisher={IEEE}
}

@article{huang2024multiobjective,
	title={Multiobjective deep reinforcement learning for computation offloading and trajectory control in {UAV}-base-station-assisted {MEC}},
	author={Huang, Hao and Chai, Zheng-Yi and Sun, Bao-Shan and Kang, Hong-Shen and Zhao, Ying-Jie},
	journal={IEEE Internet of Things Journal},
	volume={11},
	number={19},
	pages={31805--31821},
	year={2024},
	publisher={IEEE}
}

@inproceedings{gan2021trajectory,
	title={Trajectory optimization and computing offloading strategy in {UAV}-assisted {MEC} system},
	author={Gan, Youhui and He, Yejun},
	booktitle={Proc. Computing, Communications and IoT Applications (ComComAp)},
	pages={132--137},
	year={2021},
	address={Shenzhen, China}
}

@article{chen2023energy,
	title={Energy efficient task offloading and resource allocation in air-ground integrated {MEC} systems: A distributed online approach},
	author={Chen, Ying and Li, Kaixin and Wu, Yuan and Huang, Jiwei and Zhao, Lian},
	journal={IEEE Transactions on Mobile Computing},
	volume={23},
	number={8},
	pages={8129--8142},
	year={2023},
	publisher={IEEE}
}

@article{li2023computation,
	title={Computation offloading and trajectory planning of multi-{UAV}-enabled {MEC}: A knowledge-assisted multiagent reinforcement learning approach},
	author={Li, Xulong and Qin, Yunhui and Huo, Jiahao and Huangfu, Wei},
	journal={IEEE Transactions on Vehicular Technology},
	volume={73},
	number={5},
	pages={7077--7088},
	year={2023},
	publisher={IEEE}
}

@article{wang2020multi,
	title={Multi-agent deep reinforcement learning-based trajectory planning for multi-{UAV} assisted mobile edge computing},
	author={Wang, Liang and Wang, Kezhi and Pan, Cunhua and Xu, Wei and Aslam, Nauman and Hanzo, Lajos},
	journal={IEEE Transactions on Cognitive Communications and Networking},
	volume={7},
	number={1},
	pages={73--84},
	year={2020},
	publisher={IEEE}
}

@article{zhao2022multi,
	title={Multi-agent deep reinforcement learning for task offloading in {UAV}-assisted mobile edge computing},
	author={Zhao, Nan and Ye, Zhiyang and Pei, Yiyang and Liang, Ying-Chang and Niyato, Dusit},
	journal={IEEE Transactions on Wireless Communications},
	volume={21},
	number={9},
	pages={6949--6960},
	year={2022},
	publisher={IEEE}
}

@article{seid2021multi,
	title={Multi-agent {DRL} for task offloading and resource allocation in multi-{UAV} enabled {IoT} edge network},
	author={Seid, Abegaz Mohammed and Boateng, Gordon Owusu and Mareri, Bruce and Sun, Guolin and Jiang, Wei},
	journal={IEEE Transactions on Network and Service Management},
	volume={18},
	number={4},
	pages={4531--4547},
	year={2021},
	publisher={IEEE}
}

@article{shao2024deep,
	title={Deep reinforcement learning-based resource management for {UAV}-assisted mobile edge computing against jamming},
	author={Shao, Ziling and Yang, Helin and Xiao, Liang and Su, Wei and Chen, Yifan and Xiong, Zehui},
	journal={IEEE Transactions on Mobile Computing},
	volume={23},
	number={12},
	pages={13358--13374},
	year={2024},
	publisher={IEEE}
}

@article{li2023robust,
	title={Robust computation offloading and trajectory optimization for multi-{UAV}-assisted {MEC}: A multiagent {DRL} approach},
	author={Li, Bin and Yang, Rongrong and Liu, Lei and Wang, Junyi and Zhang, Ning and Dong, Mianxiong},
	journal={IEEE Internet of Things Journal},
	volume={11},
	number={3},
	pages={4775--4786},
	year={2023},
	publisher={IEEE}
}

@article{wu2025dynamic,
	title={Dynamic task allocation for {UAV} swarms in maritime rescue scenarios based on {PG-MAPPO}},
	author={Wu, Xiang and Yan, Qingzhong and Wang, Jiacun and Zhou, Yuhang and Huang, Qilong and Jiang, Changhui},
	journal={IEEE Internet of Things Journal},
	year={2025},
	volume={12},
	number={18},
	pages={38073-38087},
	publisher={IEEE}
}

@article{jiang2025dro,
	title={{DRO}-Based Computation Offloading and Trajectory Design for Low-Altitude Networks},
	author={Jiang, Guanwang and Jia, Ziye and Cui, Can and He, Lijun and Zhu, Qiuming and Wu, Qihui},
	journal={arXiv preprint arXiv:2510.23202},
	year={2025}
}

@article{wang2025joint,
	title={Joint positioning and computation offloading in multi-{UAV} {MEC} for low latency applications: A proximal policy optimization approach},
	author={Wang, Yuhui and Farooq, Junaid and Ghazzai, Hakim and Setti, Gianluca},
	journal={IEEE Transactions on Mobile Computing},
	year={2025},
	volume={24},
	number={10},
	pages={9584-9598},
	publisher={IEEE}
}

@article{zeng2019energy,
	title={Energy minimization for wireless communication with rotary-wing {UAV}},
	author={Zeng, Yong and Xu, Jie and Zhang, Rui},
	journal={IEEE Transactions on Wireless Communications},
	volume={18},
	number={4},
	pages={2329--2345},
	year={2019},
	publisher={IEEE}
}

@article{zhou2025novel,
	title={A novel scheme for dynamic task collaboration among large-scale {UAV} swarms in the era of low-altitude economy},
	author={Zhou, Jianping and Yi, Yinxue and Hu, Xiang and Deng, Xianjun and Yang, Laurence T and Zhang, Zufan},
	journal={IEEE Transactions on Cognitive Communications and Networking},
	year={2025},
	volume={12},
	pages={3117-3132},
	publisher={IEEE}
}

@article{stanimirovic2011linear,
	title={On the linear weighted sum method for multi-objective optimization},
	author={Stanimirovic, Ivan P and Zlatanovic, Milan Lj and Petkovic, Marko D},
	journal={Facta Acta Univ},
	volume={26},
	number={4},
	pages={49--63},
	year={2011}
}

@inproceedings{alsyouf2017multi,
	title={A multi-objective optimization of maintenance policies using weighted comprehensive criterion method ({WCCM})},
	author={Alsyouf, Imad and Hamdan, Sadeque},
	booktitle={Proc. IEEE ICMSAO},
	pages={1--4},
	year={2017},
	address={Sharjah, United Arab Emirates}
}

@article{liu2024two,
	title={Two-timescale synchronization and migration for digital twin networks: A multi-agent deep reinforcement learning approach},
	author={Liu, Wenshuai and Fu, Yaru and Guo, Yongna and Wang, Fu Lee and Sun, Wen and Zhang, Yan},
	journal={IEEE Transactions on Wireless Communications},
	volume={23},
	number={11},
	pages={17294--17309},
	year={2024},
	publisher={IEEE}
}

@article{shi2022joint,
	title={Joint online optimization of data sampling rate and preprocessing mode for edge--cloud collaboration-enabled industrial {IoT}},
	author={Shi, You and Yi, Changyan and Chen, Bing and Yang, Chenze and Zhu, Kun and Cai, Jun},
	journal={IEEE Internet of Things Journal},
	volume={9},
	number={17},
	pages={16402--16417},
	year={2022},
	publisher={IEEE}
}

@article{sarah2024drl,
	title={{DRL}-based availability-aware migration of a {MEC} service},
	author={Sarah, Annisa and Nencioni, Gianfranco and Khan, Md Muhidul Islam},
	journal={IEEE Open Journal of the Communications Society},
	volume={5},
	pages={5088--5102},
	year={2024},
	publisher={IEEE}
}

@article{kuang2023joint,
	title={Joint task offloading scheduling and resource allocation in air--ground cooperation {UAV}-enabled mobile edge computing},
	author={Kuang, Zhufang and Pan, Yihui and Yang, Fan and Zhang, Yan},
	journal={IEEE Transactions on Vehicular Technology},
	volume={73},
	number={4},
	pages={5796--5807},
	year={2023},
	publisher={IEEE}
}

@inproceedings{zhang2024energy,
	title={Energy-efficient task offloading in {UAV}-{RIS}-assisted mobile edge computing with {NOMA}},
	author={Zhang, Mingyang and Su, Zhou and Xu, Qichao and Qi, Yihao and Fang, Dongfeng},
	booktitle={Proc. IEEE INFOCOM WKSHPS},
	pages={01--06},
	year={2024},
	address={Vancouver, Canada}
}

@article{chen2025joint,
	title={Joint trajectory optimization and resource allocation in {UAV}-{MEC} systems: A Lyapunov-assisted {DRL} approach},
	author={Chen, Ying and Yang, Yaozong and Wu, Yuan and Huang, Jiwei and Zhao, Lian},
	journal={IEEE Transactions on Services Computing},
	year={2025},
	volume={18},
	number={2},
	pages={854-867},
	publisher={IEEE}
}

@article{zhang2025secure,
	title={Secure Task Offloading for Multi-Layer Aerial-Offshore {MEC} Networks With Imperfect Eavesdropper Information},
	author={Zhang, Yue and Na, Zhenyu and Du, Xueqi and Lin, Yun and Lin, Bin},
	journal={IEEE Wireless Communications Letters},
	year={2025},
	volume={14},
	number={11},
	pages={3814-3818},
	publisher={IEEE}
}

@article{bi2021lyapunov,
	title={Lyapunov-guided deep reinforcement learning for stable online computation offloading in mobile-edge computing networks},
	author={Bi, Suzhi and Huang, Liang and Wang, Hui and Zhang, Ying-Jun Angela},
	journal={IEEE Transactions on Wireless Communications},
	volume={20},
	number={11},
	pages={7519--7537},
	year={2021},
	publisher={IEEE}
}

@article{long2025lyapunov,
	title={Lyapunov-guided deep reinforcement learning for semantic-aware aoi minimization in {UAV}-assisted wireless networks},
	author={Long, Yusi and Gong, Shimin and Sun, Sumei and Lee, Gary CF and Li, Lanhua and Niyato, Dusit},
	journal={IEEE Transactions on Wireless Communications},
	year={2025},
	  volume={24},
	number={8},
	pages={6351-6364},
	publisher={IEEE}
}

@article{qiu2025tom,
	title={{TOM}: Joint Trajectory, Offloading and Migration Optimization in Stateful Service-Oriented {UAV}-Enabled {VEC} System},
	author={Qiu, Qijie and Li, Lingjie and Xiao, Zhijiao and Lin, Qiuzhen and Ma, Lijia and Ming, Zhong},
	journal={IEEE Transactions on Services Computing},
	year={2025},
	 volume={18},
	number={6},
	pages={4261-4275},
	publisher={IEEE}
}

@inproceedings{guo2025cognitive,
	title={Cognitive {UAV}-Assisted Offloading for Mobile Edge Computing Based on Multi-Agent Deep Reinforcement Learning},
	author={Guo, Shaoai and Li, Zan and Yan, Junjie and Zhao, Xiaohui},
	booktitle={Proc. IEEE WCNC},
	pages={1--6},
	year={2025},
	address={Milan, Italy}
}

@inproceedings{sun2024two,
	title={A two time-scale joint optimization approach for {UAV}-assisted {MEC}},
	author={Sun, Zemin and Sun, Geng and He, Long and Mei, Fang and Liang, Shuang and Liu, Yanheng},
	booktitle={Proc. IEEE INFOCOM},
	pages={91--100},
	year={2024},
	address={Vancouver, Canada}
}

@article{wang2026two,
	title={Two-Layer Reinforcement Learning-Assisted Joint Beamforming and Trajectory Optimization for Multi-{UAV} Downlink Communications},
	author={Wang, Ruiqi and Ghoura, Essra M and Alhussein, Omar and Yang, Yuzhi and Ren, Jing and Xu, Shizhong and Muhaidat, Sami},
	journal={arXiv preprint arXiv:2601.12659},
	year={2026}
}

@article{chen2022joint,
	title={Joint computation offloading and deployment optimization in multi-{UAV}-enabled {MEC} systems},
	author={Chen, Zheyi and Zheng, Hongqiang and Zhang, Jianshan and Zheng, Xianghan and Rong, Chunming},
	journal={Peer-to-Peer Networking and Applications},
	volume={15},
	number={1},
	pages={194--205},
	year={2022},
	publisher={Springer}
}

@article{xu2022cellular,
	title={Cellular-connected multi-{UAV} {MEC} networks: An online stochastic optimization approach},
	author={Xu, Yu and Zhang, Tiankui and Liu, Yuanwei and Yang, Dingcheng and Xiao, Lin and Tao, Meixia},
	journal={IEEE Transactions on Communications},
	volume={70},
	number={10},
	pages={6630--6647},
	year={2022},
	publisher={IEEE}
}

@inproceedings{guan2023mappo,
	title={{MAPPO}-based cooperative UAV trajectory design with long-range emergency communications in disaster areas},
	author={Guan, Yue and Zou, Sai and Li, Kai and Ni, Wei and Wu, Bochun},
	booktitle={Proc. IEEE WoWMoM},
	pages={376--381},
	year={2023},
	address={Boston, USA}
}

@inproceedings{liu2024heterogeneous,
	title={Heterogeneous Multi-{UAV} Task Allocation Based on Improved Multi-Objective Grey Wolf Optimization Algorithm},
	author={Liu, Senlin and Luo, Rui and Wen, Liangdong and Zhen, Ziyang and He, Jialu and Li, Yifan},
	booktitle={Proc. IEEE ICUS},
	pages={701--706},
	year={2024},
	address={Nanjing, China}
}

@article{chen2026primal,
	title={A primal-dual approach to constrained Markov decision processes with applications to queue scheduling and inventory management},
	author={Chen, Yi and Dong, Jing and Wang, Zhaoran and Zhang, Chuheng},
	journal={Management Science},
	volume={72},
	number={2},
	pages={955--988},
	year={2026},
	publisher={INFORMS}
}

@article{ding2020natural,
	title={Natural policy gradient primal-dual method for constrained markov decision processes},
	author={Ding, Dongsheng and Zhang, Kaiqing and Basar, Tamer and Jovanovic, Mihailo},
	journal={Advances in Neural Information Processing Systems},
	volume={33},
	pages={8378--8390},
	year={2020}
}

@ARTICLE{talukdar2025ddpg,
	author={Talukdar, Nayanjit and Raghav, Aashish and Hazra, Abhishek and Barman, Dipankar Ch. and Mazumdar, Nabajyoti},
	journal={IEEE Internet of Things Journal},
	title={A Deep Deterministic Policy Gradient Method for Optimizing Task Completion Time and Energy Efficiency in {UAV}-Assisted {IoT} Networks},
	year={2025},
	volume={12},
	number={15},
	pages={31907--31917},
	doi={10.1109/JIOT.2025.3575714}
}

@ARTICLE{talukdar2026mappo,
	author={Talukdar, Nayanjit and Barai, Tushar and Hazra, Abhishek and Mazumdar, Nabajyoti},
	journal={IEEE Transactions on Sustainable Computing},
	title={{MAPPO}-Driven Task Quality Optimization With Wireless Energy Transfer in Multi-{UAV} {MEC}-Enabled {IoT} Networks},
	year={2026},
	volume={11},
	number={3},
	pages={302--315},
	doi={10.1109/TSUSC.2026.3688744}
}

@article{zeng2017energy,
	title={Energy-efficient {UAV} communication with trajectory optimization},
	author={Zeng, Yong and Zhang, Rui},
	journal={IEEE Transactions on Wireless Communications},
	volume={16},
	number={6},
	pages={3747--3760},
	year={2017},
	publisher={IEEE}
}

@article{al2014optimal,
	title={Optimal {LAP} altitude for maximum coverage},
	author={Al-Hourani, Akram and Kandeepan, Sithamparanathan and Lardner, Simon},
	journal={IEEE Wireless Communications Letters},
	volume={3},
	number={6},
	pages={569--572},
	year={2014},
	publisher={IEEE}
}

@article{friis1946note,
	title={A note on a simple transmission formula},
	author={Friis, Harald T},
	journal={Proceedings of the IRE},
	volume={34},
	number={5},
	pages={254--256},
	year={1946},
	publisher={IEEE}
}

\end{document}